\pdfoutput=1
\documentclass[pra,aps,showpacs,unsortedaddress,superscriptaddress,twocolumn]{revtex4}
\usepackage{graphicx}
\usepackage{float}

\begin{document}

\title{Intrinsic phonon effects on analog quantum simulators with ultracold trapped ions
}

\author{C.-C. Joseph Wang}
\email{joseph@physics.georgetown.edu}
\affiliation{Department of Physics, Georgetown University, 37th and O Sts. NW, Washington, DC 20057, USA}
\author{J.~K.~Freericks}
\affiliation{Department of Physics, Georgetown University, 37th and O Sts. NW, Washington, DC 20057, USA}

\date{\today}


% definitions
\def\bx{{\bf x}}
\def\bk{{\bf k}}
\def\br{{\bf r}}
\def\bu{{\bf u}}
\def\half{\frac{1}{2}}
\def\args{(\bx,t)}

\begin{abstract}
 Linear Paul traps have been used recently to simulate the transverse field Ising model with long-range spin-spin couplings.
We study the intrinsic effects of phonon creation (from the initial phonon ground state) on the spin-state probability and spin entanglement for such quantum spin simulators. While it has often been assumed that phonon effects are benign because they play no role in the pure Ising model, they can play a significant role when a transverse field is added to the model. We use a many-body factorization of the quantum time-evolution operator of the system, adiabatic perturbation theory and exact numerical integration of the Schr\"odinger equation in a truncated spin-phonon Hilbert space followed by a tracing out of the phonon degrees of freedom to study this problem. We find
that moderate phonon creation often makes the probabilities of different spin states behave differently from the static spin Hamiltonian. In circumstances
in which phonon creation is minor, the spin dynamics state probabilities converge to the static spin Hamiltonian prediction at the cost of reducing the spin entanglement.  We show how phonon creation can severely impede the observation of kink transitions in frustrated spin systems when the number of ions increases. Many of our results also have implications
for quantum simulation in a Penning trap.
\end{abstract}
\maketitle
\section{Introduction}

Complex states of matter like spin liquids are suspected to
exist in quantum spin models with frustration due to geometry or due to the nature of the spin-spin interaction~\cite{spinliquid_Balents,spinliquid_Fisher1,spinliquid_Varney}. Spin liquids are complicated quantum many-body states that exhibit significant entanglement  of their wave functions without symmetry breaking, and could also exhibit emergent quantum phenomena within their low-energy excitation spectra. Classical computation, such as exact diagonization and quantum Monte Carlo simulation, or conventional theories based on local order parameters fail to describe these systems without bias.  For example, exact diagonalization studies are limited to small size lattices and hence usually have strong finite-size effects, while quantum Monte Carlo simulations can suffer from the sign problem or have a large computational expense to describe long-range interactions and hence cannot reach the low temperatures needed to see the predicted exotic phases.

Feynman proposed that one could use controlled quantum-mechanical systems with few quantum gates to simulate many-body problems~\cite{feynman,Seth} as an useful quantum computation before achieving universal quantum computation. In recent years, there has been significant success in trying to achieve this goal by quantum simulation of desired spin models through analogous cold atom systems~\cite{emulators,emulators1,emulators2}. We focus here on one platform for performing analog quantum computation, the simulation of interacting quantum spins via manipulation of hyperfine states of ions in a linear Paul trap~\cite{gate,three-ion,Kim,Islam,A.Friedenauer}
although many ideas presented here can be generalized to adiabatic
quantum state computation in the two dimensional Penning trap as well~\cite{emulators2}.  In the Paul trap systems, clock states of the ions (states with no net $z$-component of angular momentum) are the pseudospin states, which can be manipulated independently by a pseudospin-dependent force driven by laser beams. The lasers couple the pseudospin states to the lattice vibrations of the trapped ions, which leads to effective spin-spin interactions when the phonon degrees of freedom are adiabatically eliminated~\cite{spin-spin-interactions,Duan,Kim} based on  the idea of geometric phase gate~\cite{Leibfried} or M{\o}lmer-S{\o}rensen gate~\cite{Sorensen}.
Theoretically, the analog ion-trap simulators can be described as nonequilibrium driven quantum systems with both spin and phonon degrees of freedom.  Sufficiently small systems can be treated numerically in an exact fashion by truncating the phonon basis and taking into account all possible quantum states in the solution of the time-dependent Schr\"odinger equation.
Experimentally, ion traps have been used to simulate the transverse-field Ising model with a small number of ions ~\cite{A.Friedenauer,three-ion,Islam} based on simulated quantum annealing~\cite{quantum annealing} (see Ref.~\onlinecite{kim_review} for a review).  It has been known experimentally that moderate phonon creation
is commonplace (on the order of one phonon per mode)~\cite{Islam}, even when the system is cooled to essentially the phonon ground state prior to the start of the simulation. In addition, the role phonons play
are intrinsic and essential for the mediated spin-spin interaction in trapped ion systems especially in the presence of noncommuting magnetic
field Hamiltonian in addition to the spin Hamiltonian of interest. Therefore, an understanding of the role phonons play in the spin simulator is crucial to understanding its accuracy.

The organization of this paper is as follows.
In Sec. II, we describe the microscopic Hamiltonian for the ion-trap-based simulators and then show how one can factorize the time-evolution operator into a pure phonon term, a coupled spin-phonon term, a pure spin-spin interaction term, and a complicated term that primarily determines the degree of entanglement of the spins. Next we use adiabatic perturbation theory to determine how adiabatic state evolution can be used to reach a complicated, potentially spin-liquid-like ground state, and detail under what circumstances the evolution is not adiabatic (diabatic).
In Sec. III, we show numerical comparison studies in various relevant circumstances  based on a direct integration of the time-dependent Schr\"odinger equation, including both spin and phonon degrees of freedom (the latter in a truncated basis).
In Sec. IV, we conclude with discussions and possible experimental limitations and improvements.

\section{Theory}
\subsection{Microscopic Hamiltonian}
When $N$ ions are placed in a linear Paul trap~\cite{three-ion,Kim,MichaelJohanning} with harmonic trapping potentials, they form a nonuniform (Wigner) lattice, with increasing interparticle spacing as one moves from the center to the edge of the chain. The ions vibrate in all three spatial dimensions about these equilibrium positions~\cite{James} with $3N$ normal modes.
Two hyperfine clock states (relatively insensitive to external magnetic field fluctuations because the $z$-component of total angular momentum is zero) in each ion will be the pseudospins (and are split by an energy difference $\hbar\omega_0$).
Hence, the bare Hamiltonian $H_{0}$ including the pseudospin and motional degrees of freedom  for the ion chain is given by
\begin{equation}
  H_{0}=\sum_{j}\frac{\hbar\omega_{0}}{2}\sigma_{j}^z+\sum_{\alpha \nu}\hbar \omega_{\alpha \nu}(a^{\dag}_{\alpha \nu} a_{\alpha \nu}+\frac12),
\end{equation}
where $\sigma_j^z$ is the Pauli spin matrix at the $j$th ion site and
the second term is the phonon Hamiltonian $H_{ph}$ with the phonon creation operator of the normal mode $\nu$ along the three spatial directions $\alpha \in {X,Y,Z}$. The notation $x,y,z$ refers to the
pseudospin orientation in the Bloch sphere.
The $\alpha$th spatial component of the $j$th ion displacement operator $\delta {\hat R}_{j}^{\alpha}$ is related to the $\alpha$th phonon normal mode amplitude (unit norm eigenvector of the dynamical matrix) $b_{j}^{\alpha \nu}$ and the $\alpha$th phonon creation and annihilation operator via $\delta {\hat R}_{j}^{\alpha}=\sum_{\nu} b_{j}^{\alpha \nu}\sqrt{\frac{\hbar}{2M\omega_{\alpha \nu}}}[a_{\alpha \nu}+a^{\dag}_{\alpha \nu}]$ with $M$ the mass of the ion and $\omega_{\alpha\nu}$ the normal-mode frequency.

 A laser-ion interaction is imposed to create
 a spin-dependent force on the ions by using bichromatic laser beams to couple these clock states to a third state via stimulated Raman transitions~\cite{PJLee}. Effectively, this process is equivalent to an off-resonant laser coupling to the two clock states by a small frequency detuning $\mu$ determined by the frequency difference of the bichromatic lasers. The ions are crystallized along the easy axis ($Z$-axis) of the trap with hard axes in the $X$- and $Y$-directions where the transverse phonons lie. Then coupling the Raman lasers in the transverse direction minimizes effects of ion heating and allows for an identical spin axis for each ion~\cite{Duan}. By accurate control of the locked phases of the blue detuned and red detuned lasers with similar Rabi frequencies, an effective laser-ion Hamiltonian~\cite{PJLee,Wineland}
along the spin direction $\sigma^x$ can be engineered in the Lamb-Dicke limit $\Delta k |\delta {\bf\hat R}_{j}(t)|\ll 1$  :
\begin{equation}
\mathcal{H}_{LI}(t)= -\hbar \sum_{j=1}^{N}\Omega_{j}^{e}\Delta {\bf k}\cdot\delta {\bf \hat R}_{j}(t)\sigma_{j}^{x} \sin(\mu t),%This form is not correct for any \phi_{s}
\label{eq:phonon}
\end{equation}
in which the effective Rabi frequency $\Omega_{j}^{e}$ is generated by one effective blue-detuned beam and one red-detuned beam simultaneously
(refer to Appendix A for details).

In experiments, one uses adiabatic quantum state evolution to evolve the ground state from an easily prepared state to the desired complex quantum state that will be studied. For spin models generated in an ion trap, it is easy to create a fully polarized ferromagnetic state along the $z$-direction via optical pumping, and then apply a spin rotation (for instance, with a pulsed laser) to reorient the ferromagnetic state in any direction (usually chosen to be the $y$-direction).  Then, if one introduces a Hamiltonian with a magnetic field in the direction of the polarized state, it is in the ground state of the system. By slowly reducing the magnitude of the field and turning on the spin Hamiltonian of interest, one can adiabatically reach the ground state of the Hamiltonian (at least  in principle).  Hence, one has an additional
Zeeman term with a spatially uniform time-dependent effective magnetic field ${\bf B}(t)$ coupled to the different Pauli spin matrices as
\begin{equation}
\mathcal{H}_{B}(t)=\sum_{j=1}^{N}{\bf B}(t)\cdot{{\bf \hat \sigma}_{j}}
\end{equation}
(the magnetic field has units of energy here, since we absorbed factors of the effective magnetic moment into the definition of {\bf B}, and it can also be expressed in units of frequency if we use units with $\hbar=1$).
Note that we are using an unconventional sign for the coupling to the magnetic field, since this is the sign convention often used in the ion-trapping community.  In this case, the ground state of the magnetic field Hamiltonian has the spins aligned opposite to that of the field, while the highest-energy state has them aligned with the field. The magnetic field ${\bf B}(t)$  is made in the $y$-direction by directly  driving a resonant radio-frequency field
with frequency $\omega_{0}$ between the two hyperfine states
to implement the spin flips~\cite{Wineland} or by indirect Raman coupling through a third state to effectively   couple the two hyperfine states~\cite{PJLee}.
The full Hamiltonian is then $\mathcal{H}(t)=\mathcal{H}_{ph}+\mathcal{H}_{LI}(t)+\mathcal{H}_{B}(t)$.

\subsection{Factorization of the time evolution operator}
We solve for the quantum dynamics of this time-dependent Hamiltonian by calculating the evolution operator as a time-ordered product
$U(t,t_0)=\mathcal{T}_t \exp[-i\int_{t_0}^t dt^\prime \mathcal{H}(t^\prime)/\hbar]$ and operating it on the initial quantum state $|\psi(t_0)\rangle$.  For the adiabatic evolution of the ground state, we start our system in a state with the spins aligned along the magnetic field and the system cooled down so that there are no phonons at time $t_0$: $|\psi(t_0)\rangle={|\uparrow_y \uparrow_y\ldots\uparrow_y\rangle}\otimes |0\rangle_{ph}$. This means we will be following the highest excited spin state of the system, as described in more detail below. While it is also possible to examine incoherent effects due to thermal phonons present at the start of the simulation, we do not do that here, and instead focus solely on intrinsic phonon creation due to the applied spin-dependent force.

 In time-dependent perturbation theory, one rewrites the evolution operator in the interaction picture with respect to the time-independent part of the Hamiltonian.  This procedure produces an exact factorized evolution operator
\begin{equation}
U(t,t_0)=e^{-\frac{i}{\hbar} \mathcal{H}_{ph}(t-t_{0})} U_{I}(t,t_{0})
\label{eq: interaction_picture}
\end{equation}
which is the first step in our factorization procedure [the first factor is called the phonon evolution operator $U_{ph}(t,t_0)$.] [Note that $\mathcal{H}_{ph}$ is time independent and it is multiplied by the factor $(t-t_0)$ in the exponent.] The second factor is the evolution operator in the interaction picture, which satisfies an equation of motion given by $i \hbar {\partial U_I(t,t_{0})}{/\partial t}=[V_I(t)+\mathcal{H}_B(t)]U_I(t,t_{0})$, with $V_I(t)=\exp[i \mathcal{H}_{ph} (t-t_0)/\hbar ]\mathcal{H}_{LI}(t)\exp[-i \mathcal{H}_{ph}(t-t_0)/\hbar]$  since $[\mathcal{H}_B(t),\mathcal{H}_{ph}]=0$. The only difference between $\mathcal{H}_{LI}(t)$ and $V_I(t)$ is that the phonon operators are replaced by their interaction picture values: $a_{\alpha\nu}\rightarrow a_{\alpha\nu}\exp [-i\omega_{\alpha\nu}(t-t_0)]$ and $a_{\alpha\nu}^\dagger
\rightarrow a_{\alpha\nu}^\dagger\exp [i\omega_{\alpha\nu}(t-t_0)]$.

We now work to factorize the evolution operator further.  Motivated by the classic problem on driven harmonic oscillators ~\cite{gottfried}, we factorize the interaction picture evolution operator via $U_{I}(t,t_{0})=\exp[-i W_{I}(t)/\hbar]\bar{U}(t,t_0)$ with $W_I(t)$ defined by $W_I(t)=\int_{t_0}^t dt^\prime V_I(t^\prime)$ (we call the factor on the left the phonon-spin evolution operator $U_{ph-sp}=\exp[-iW_I(t)/\hbar]$ and the one on the right is the remaining evolution operator).  The key step in this derivation is that the multiple commutator satisfies $[[W_I(t),V_I(t^\prime)],V_I(t^{\prime\prime})]=0$.  This fact greatly simplifies the analysis below.

The equation of motion for the remaining evolution operator $\bar U(t,t_{0})$ satisfies
\begin{equation}
i\hbar \frac{\partial}{\partial t}{\bar U}(t, t_{0})={\bar \mathcal{H}}(t){\bar U}(t, t_{0}),
\end{equation}
in which the operator ${\bar \mathcal{H}}(t)$ is given by the expression
\begin{equation}
{\bar \mathcal{H}}(t)=e^{\frac{i}{\hbar}W_{I}(t)}[-i\hbar \partial_{t}+\mathcal{H}_{B}(t)+V_{I}(t)]e^{-\frac{i}{\hbar}W_{I}(t)}.
\end{equation}
The operator ${\bar \mathcal{H}}(t)$ can then be expanded order by order as
\begin{eqnarray}
\label{eq:main}
e^{\frac{i}{\hbar}W_{I}(t)}V_{I}(t)e^{-\frac{i}{\hbar}W_{I}(t)}&=& V_{I}(t)+
\frac{i}{\hbar}\left[W_{I}(t), V_{I}(t)\right] \\
\label{eq:main_b}
e^{\frac{i}{\hbar}W_{I}(t)}\mathcal{H}_{B}(t)e^{-\frac{i}{\hbar}W_{I}(t)} &=&  \sum_{j=1}^{N}\Big\{ {\bf B}(t)\cdot\hat \sigma_{j}+\\ &&\stackrel{\underbrace{\frac{i}{\hbar}
[W_{I}(t),{\bf B}(t)\cdot\hat \sigma_{j}]
+ \ldots}\Big\}}{\rm{Residual \ terms}} , \nonumber \\
e^{\frac{i}{\hbar}W_{I}(t)}i\hbar\partial_{t}e^{-\frac{i}{\hbar}W_{I}(t)}&=&  V_{I}(t)\nonumber\\
&+&\frac{1}{2}\frac{i}{\hbar}[W_{I}(t),V_{I}(t)]
\end{eqnarray}
where we used the facts that $\partial{W_{I}}(t)/\partial t=V_{I}(t)$ and $[[W_I(t),V_I(t^\prime)],V_I(t^{\prime\prime})]=0$.
%%%%%%%%%%%%%%%%%%%%%%%%%%%%%%%%%%%%%%%%
Explicit calculations then yield
\begin{eqnarray}
V_{I}(t) &=& - \sum_{j=1}^{N}\sum_{\nu=1}^{N}\sum_{\alpha}\hbar\Omega_{j}\eta_{\alpha \nu} b_{j}^{\alpha \nu}(a_{\alpha \nu}e^{-i\omega_{\alpha \nu}t}+
a^{\dag}_{\alpha \nu} e^{i\omega_{\alpha \nu}t})\nonumber \\
&\times&\sin(\mu t)\sigma_{j}^{x},
%W_{I}(t)&=&-\sum_{j=1}^{N}\sum_{\nu=1}^{N}\sum_{\alpha}\frac{\hbar\Omega_{j}\eta_{\alpha \nu}b_{j}^{\alpha \nu}}{\omega_{\alpha \nu}^{2}-\mu^2}\sigma_{j}^{x}
%&\times&
%\Big \{ \Big [ e^{-i\omega_{\alpha \nu}t}(i\omega_{\alpha \nu} \sin{\mu t}+\mu\cos{\mu t})\nonumber
%&-&  e^{-i\omega_{\alpha \nu}t_{0}}(i\omega_{\alpha \nu} \sin{\mu %t_{0}}+\mu\cos{\mu t_{0}}) \Big ] a_{\alpha \nu} +  h. c. \Big \}, %\nonumber
\end{eqnarray}
in which $\eta_{\alpha \nu}=\delta_{\alpha X}\delta k^{\alpha}\sqrt{\hbar/2M\omega_{\alpha \nu}}$ is the Lamb-Dicke parameter for the phonon mode $\alpha \nu$ with the $\alpha$th component of the laser momentum  $\delta k^{\alpha}$. When the terms in Eq.~(\ref{eq:main_b}) vanish, virtually excited phonons will be shown to play no role on the spin-state probabilities as a function of time, but in the presence of a transverse field, due to the noncommuting nature of quantum operators, phonon creation can significantly affect the spin-state probabilities. This fact has not been considered in detail before, and involves one of the most important results of our work, as detailed below.

\subsection{Ising spin models and M{\o}lmer-S{\o}rensen gate for vanishing transverse magnetic field}
 With a vanishing transverse magnetic field, the Hamiltonian $\bar H$
 can be greatly reduced to the spin-only Hamiltonian $H_{spin}(t)=\frac{i}{2\hbar}[W_{I}(t),V_{I}(t)]$. Because the spin operators $s_{j}^{x}$ in $W_{I}$ and $V_{I}$ commute, one can exactly derive the following Ising spin Hamiltonian~\cite{spin-spin-interactions,Duan}
 \begin{equation}
 H_{spin}(t)=\sum_{j,j'=1}^N J_{jj'}(t')\sigma_{j}^{x}\sigma_{j'}^{x}.
 \end{equation}
Then the expression for the spin exchange interaction $J_{jj'}(t)$ is
\begin{widetext}
\begin{equation}
\label{eq:exchange}
 J_{jj'}(t)=\frac{\hbar}{2}\sum_{\nu=1}^{N}\frac{\Omega_{j}\Omega_{j'}\eta_{X \nu}^{2}b_{j}^{X \nu}b_{j'}^{X \nu}}{\mu^2-\omega_{X \nu}^2}[\omega_{X\nu}-\omega_{X \nu}\cos{2\mu t}-2\mu\sin{\omega_{X \nu} t}\sin{\mu t}],
 \end{equation}
\end{widetext}
which can be uniformly antiferromagnetic  ($J_{jj'}(t) > 0$) or ferromagnetic ($J_{jj'}(t) < 0$) for the instantaneous ground state of the Hamiltonian $H_{spin}(t)$
 when the laser detuning $\mu$ is detuned close to center of mass mode frequency $\omega_{CM}$.
However, the interaction $J_{jj'}(t) > 0$ can also be
inhomogeneous and frustrated  when the laser is detuned in between phonon modes with details depending on the properties of the nearby
phonon modes $\omega_{X\nu}$, $b_{j}^{X\nu}$, and $b_{j'}^{X\nu}$.

The M{\o}lmer-S{\o}rensen gate~\cite{Sorensen} was originally proposed to disentangle phonon effects from the spins in ion-trap quantum computing. It was discovered that because the phonons are harmonic, one could operate on the spins in such a way that the phonon state is unmodified after the gate operation (irrespective of the initial population of phonons). But one needs to keep in mind that this gate has no transverse field present, which can modify it because the transverse field operator does not commute with the Ising Hamiltonian.
 We begin our discussion by assuming that the laser is closely detuned to the transverse center of mass mode with angular frequency $\omega_{CM}$ ($|\mu-\omega_{CM}|\ll \mu$) and the addressing laser intensity for each ion is uniform and moderate ($\Omega_{j}=\Omega \ll \mu$). In this situation, the time dependent term with the frequency $\mu+\omega_{CM}$
 in the interaction $V_{I}$ can be neglected. Therefore, the interaction
$V_{I}(t)$ and the operator $W_{I}(t)$, which are proportional to the collective spin operator $S_{x}=\sum_{j=1}^{N}s_{j}^{x}$, can be reduced to the following forms
\begin{eqnarray}
      V_{I}(t) &=& i\frac{\hbar\eta_{CM}\Omega}{2\sqrt{N}}S_{x} \big[
      e^{-i(\omega_{CM}-\mu)t}a_{CM}+ h.c.\big], \\
      W_{I}(t) &=& \frac{\hbar \eta_{CM}\Omega }{2\sqrt{N}(\Omega_{CM}-\mu)}S_{x}  \nonumber \\
      &\times & {\big[(1-e^{-i(\omega_{CM}-\mu)t})a_{CM} +h. c.\big]}
      \label{eq:Molmer}
\end{eqnarray}
where $\eta_{CM}$ is the Lamb-Dicke parameter for the center of mass mode and $t_{0}=0$ is chosen.

There are two parameter regimes where phonon effects disappear.
In the weak-coupling regime $\eta_{CM} \Omega \ll \mu-\omega_{CM}$, the operator $W_{I}(t)$ almost vanishes and the time evolution operator $U_{I}$ is solely determined by the spin-only Hamiltonian because the phonon dynamics are adiabatically eliminated. Any extra phonon state redistribution takes a long time to be experimentally observable and therefore phonon
effects are under control. Outside the weak-coupling regime, one can also prevent phonon effects  by preparing spin states determined by the spin Hamiltonian $H_{spin}(t)$ at a particular waiting time interval $t=T$. The idea is to choose the time interval $T$ such that the operator is periodic with
integer $K$ cycles so that $W_{I}(KT)=W_{I}(0)$. Therefore, the initial phonon state at the start of the simulation will be revived at the time intervals $(\omega_{CM}-\mu)T=K\times 2\pi$ as can be clearly seen from Eq.~(\ref{eq:Molmer}) when the we start with the phonon ground state $|0\rangle$ for example, but it is generally true for any occupancy of the phonon states.

\subsection{Effective spin Hamiltonian with a transverse field}
In the presence of a transverse magnetic field, the ideas of the M{\o}lmer-S{\o}rensen gate are modified.
While it is tempting to claim that the residual spin-phonon terms in the magnetic field are irrelevant in the
Lamb-Dicke limit  $\eta_{\alpha \nu} \ll 1$, it is difficult to quantify this if the residual terms are relevant in the presence of the time
dependent magnetic field ${\bf B}(t)\cdot\hat \sigma_{j}$ which can be large in magnitude. In fact, phonon effects often modify the time evolution of the spin states when a transverse field is present.
However, one can say that in cases where the integral of the field over time is small (which occurs when the field is small, or when it is rapidly ramped to zero) or when the ${\bf B}$ field lies along the $x$-direction only (or vanishes), the  residual terms in Eq.~(\ref{eq:main_b}) are irrelevant.
For large detuning (weak-coupling regime), where $|W_{I}(t)|/\hbar \ll 1$ or ${\eta_{X\nu}\Omega_{j}} \ll {\omega_{X\nu}-\mu}$,
the residual terms are always higher-order perturbations with respect to the leading transverse magnetic field term ${\bf B}(t)\cdot\hat \sigma_{j}$ in the course of the quantum simulation.
On those occasions, one  can also consider the system as described by only the quantum Ising spin model in a transverse magnetic field.  In general though, we need to determine how large the residual terms are, which often can only be done with numerical calculations. We illustrate this for a number of different cases below.
The residual terms are
\begin{equation}
\mathcal{H}_{res}(t)=\exp[iW_I(t)/\hbar]\mathcal{H}_B(t)\exp[-iW_I(t)/\hbar]-\mathcal{H}_B(t).
\end{equation}
The equation of motion for $\bar U$ can be written as
\begin{eqnarray}
i\hbar\frac{\partial}{\partial t}\bar U(t,t_0)&=&\left \{\frac{i}{2\hbar}[W_I(t),V_I(t)]+\mathcal{H}_B(t)+\mathcal{H}_{res}(t)\right \}\nonumber\\
&\times&\bar U(t,t_0).
\end{eqnarray}
We perform the final factorization by writing $\bar U(t,t_0)=U_{spin}(t,t_0)U_{ent}(t,t_0)$ where the spin-evolution operator satisfies $i\hbar\partial_t U_{spin}(t,t_0)=\{i[W_I(t),V_I(t)]/2\hbar+\mathcal{H}_B(t)\}U_{spin}(t,t_0)$ and the entangled evolution operator satisfies
\begin{equation}
i\hbar\frac{\partial}{\partial t} U_{ent}(t,t_0)=U^\dagger_{spin}(t,t_0)\mathcal{H}_{res}(t)U_{spin}(t,t_0) U_{ent}(t,t_0).
\end{equation}
The spin evolution operator ${U}_{spin}(t,t_{0})$ becomes
\begin{widetext}
\begin{equation}
\label{eq:Hspin}
{ U}_{spin}(t,t_{0})  =  \mathcal{T}_{t}\exp\left[-\frac{i}{\hbar}\int_{t_{0}}^{t}dt' \left(\sum_{j,j'=1}^N J_{jj'}(t')\sigma_{j}^{x}\sigma_{j'}^{x}+
B(t')\sum_{j=1}^N\sigma_{j}^{y} \right)\right]=
\mathcal{T}_{t}\exp\left[-\frac{i}{\hbar}\int_{t_{0}}^{t}dt' H_{spin}(t')\right]
,
\end{equation}
\end{widetext}
which is the third factor for the evolution operator of the Ising model in a transverse field and we define $H_{spin}(t)$ in the exponent.  The spin exchange terms  $J_{jj'}(t)=J_{jj'}^{0}+\Delta J_{jj'}(t)$ as given in Eq.~(\ref{eq:exchange}) include a time-independent exchange interaction between two ions $J_{jj'}^{0}$  and a time-dependent exchange interaction $\Delta J_{jj'}(t)$. The time-independent term can be thought of as the effective static spin-spin Hamiltonian that is being simulated, while the time-dependent terms can be thought of as diabatic corrections, which are often small in current experimental set-ups, but need not be neglected. For simplicity, we set the initial time $t_{0} = 0$.

The entanglement evolution operator $U_{ent}$ is a complicated object in general, but it simplifies when one can  approximate the operator $U_{ent}$ as $U_{ent}\approx 1$ for the special situations discussed at the end of the last subsection.  In general, this evolution operator involves a coupling of spins to phonons in all directions and has a very complicated time dependence. If one evaluates the first few terms of the series for the time-ordered product, one finds it involves multispin interactions, spin-phonon coupling, and spin-exchange interactions in all spatial directions. But the net weight of all of the terms is governed by the integral of the magnetic field over time, so if that integral is small, then this factor will also be small. Therefore, the adiabatic elimination of phonons based on M{\o}lmer-S{\o}rensen gate~\cite{Sorensen} can be justified only in the case of a vanishing transverse magnetic field. With a constant magnetic field, the entanglement between spins and phonons can be periodic so that phonon effects can continue to be nulled at integer multiples of the appropriate period~\cite{Leibfried,Sorensen}. But such a procedure would be more complicated than the standard gate, and is not relevant for adiabatic state creation simulations, so we won't discuss it further here.
From a mathematical standpoint, because the entanglement evolution operator is on the far right of the factorization, it's main effect is to modify the state from an initial spin state in direct product with the phonon vacuum to a state that will typically involve some degree of entanglement between phonon and spin degrees of freedom.

We can use this factorization to show that in cases where the spin-entanglement evolution operator can be approximated by the unit operator, then  phonons have no observable effects on the probability of product states (regardless of the number of coherent phonons created during the simulation), so this result is similar in spirit to the original M{\o}lmer-S{\o}rensen gate, but is different because it holds in the presence of a transverse field and requires no special times for periodic variations to recur.  To do this, we need one final identity. We further factorize the entangled phonon-spin evolution operator $\exp [-i W_{I}(t)/\hbar ]$ into
the product $\exp [-i\sum_{\nu}{\Gamma}_{X \nu}^{*}(t)a^{\dag}_{x \nu}]$ $\exp[-i\sum_{\nu} \Gamma_{X \nu}(t)a_{x \nu}]\exp[-(1/2)\sum_{\nu}\Gamma_{X \nu}(t){\Gamma}_{X \nu}^{*} (t)]$
with the spin operator defined to be ${\Gamma}_{X \nu}^{*}(t)=\sum_{j}\gamma_{X j}^{\nu}(t)\sigma_{j}^{x}$
and its complex conjugate is $\Gamma_{X \nu}(t)$, while the function $\gamma_{X j}^{\nu}(t)$ satisfies
$\gamma_{X j}^{\nu}(t)=\Omega_{j}\eta_{X \nu}b_{X j}^{\nu}/(\mu^2-\omega_{X \nu}^{2})\times[\exp\{-i\omega_{X \nu}t\}(i\omega_{X \nu}\sin\mu t+\mu \cos\mu t)-\mu]$.

At this stage, we have factorized the evolution operator into four main terms, each term being an evolution operator evolving the system from time $t_0$ to time $t$. We have explicit values for the first three factors, but the last term (the entanglement evolution operator) can be quite complicated; we have also described situations where the exponent of that term is small and can be neglected. In this case, the probabilities to observe any of the $2^{N}$ product states with a quantization axis along the Ising axis ($|\beta \rangle = |\uparrow_{x} \text{or} \downarrow_{x}\rangle \otimes |\uparrow_{x} \text{or} \downarrow_{x}\rangle\otimes ......$ for the $N$ ionic spins) is unaffected by the presence of an arbitrary number of real excited phonons (which are excited by the phonon-spin evolution operator). Using the fundamental axiom of quantum mechanics, the probability $P_\beta(t)$ to observe a product spin state $|\beta \rangle$ starting initially from the phonon ground state $|0\rangle$ and not measuring any of the final phonon states involves the trace over all possible final phonon configurations
\begin{widetext}
\begin{equation}
\label{eq:prob}
P_\beta(t)=\sum_{n_{X1}=0}^{\infty}...\sum_{n_{X\nu}=0}^{\infty}...\sum_{n_{X N}=0}^{\infty}|\langle\beta| \otimes\langle n_{X 1}.. n_{X \nu}..n_{X N}|  U(t,t_{0})|0\rangle \otimes |\Phi(0)\rangle|^{2}
\end{equation}
\end{widetext}
where $|\Phi(0)\rangle$ is any initial spin state (it need not be a product state) and $n_{X\nu}$ denotes the number of phonons excited in the $\nu$th mode in the $X$-direction. The operator in the matrix element entangles the phonons and the spins, so we evaluate the matrix element in two steps: (1) first we evaluate the phonon part of the operator expectation value, and then (2) we evaluate the spin part.  Note that since the pure phonon factor of the phonon evolution operator $\exp[-i\mathcal{H}_{ph}t/\hbar ]$ is a phase factor, it has no effect on the probabilities when evaluated in the phonon number operator basis, so we can drop that factor. Next, the term $\exp[-i\sum_{\nu} \Gamma_{X \nu}(t)a_{X \nu}]$ gives 1 when operating on the phonon vacuum to the right, so it can be dropped. We are thus left with three factors in the evolution operator.  One involves exponentials of the phonon creation operator multiplied by spin operators (and is essentially a coherent-state excitation for the phonons with the average phonon excitation number determined by the spin state being measured), one involves products of spin operators that resulted from the factorization of the coupled phonon-spin evolution operator factor, and one is the pure spin evolution factor $U_{spin}$. The two remaining factors that appear on the left involve only $\sigma_{j}^{x}$ spin operators, and hence the product state basis is an eigenbasis for those operators.  This fact allows us to directly evaluate the expression in Eq.~(\ref{eq:prob}). We expand the evolution of the initial state at time $t$ in terms of the product-state basis
$|\Phi(t)\rangle = {U_{spin}}(t,t_{0}=0)|\Phi (0)\rangle=\sum_{\beta^\prime} c_{\beta^\prime}(t)|{\beta^\prime}\rangle$ with $|\beta^\prime\rangle$ denoting each of the $2^N$ product state basis vectors and $c_{\beta^\prime}(t)$ is a (complex) number.
Using the fact that the product states satisfy the eigenvalue equation $\sigma_{j}^{x}|\beta^\prime\rangle =
m_{j\beta^\prime}^{x}|\beta^\prime\rangle$ with eigenvalues $m_{j\beta^\prime}^{x}=+1 $ (for $|\uparrow_{x}\rangle$) or $-1$ (for $|\downarrow_{x}\rangle$), we arrive at the expression for the probability
\[
P_{\beta}(t) = |c_{\beta}(t)|^2 \exp{\left [ -\sum_{\nu j j'}\gamma^{\nu}_{Xj}(t)\gamma^{*\nu}_{X j'}(t)m_{j\beta}^{x}m_{j'\beta}^{x}\right ]}
\]
\begin{equation}
\times
\sum_{n_{X1}...n_{X N}=0}^{\infty}
\prod_{\nu=1}^N\left [ \frac{1}{n_{X\nu}!}\left \{\sum_{jj'}\gamma_{Xj}^{\nu}(t)\gamma_{Xj'}^{*\nu}(t)
m_{j\beta}^{x}  m_{j'\beta}^{x} \right\}^{n_{X\nu}}\right ] .
\end{equation}
We used the matrix element
\[
\langle n_{X1}.. n_{X\nu}..n_{XN}|e^{-i\sum_{\nu}{\Gamma}_{X \nu}^{*}(t)a^{\dag}_{X\nu}}|0\rangle
\]
\begin{equation}
=\frac{{(-i)}^{n_{X1}+...n_{X\nu}+..n_{XN}}}{\sqrt{n_{X1}!..n_{X\nu}!..n_{XN}!}}{\Gamma}_{X 1}^{*}(t)^{n_{X1}}...{\Gamma}_{X \nu}^{*}(t)^{n_{X\nu}}
...{\Gamma}_{X N}^{*}(t)^{n_{XN}}
\end{equation}
in the derivation. The summations become exponentials, which exactly cancel the remaining exponential term and finally yield $P_{\beta}(t)=|c_{\beta}(t)|^2$, which is what we would have found if we evaluated the evolution of the spins using just the spin evolution operator $U_{spin}$ and ignoring the phonons altogether.
Hence, \textit{the coherently excited phonons have no observable effects on the probability of product states for the transverse-field quantum Ising model when we can neglect the entanglement evolution operator.}
If we do not measure the probability of product states, then the terms from the coupled spin-phonon evolution operator remain spin operators, and one can show that the probabilities are changed by the phonons.  In other words, it is because the spin-phonon evolution operator is diagonal in the product space basis for phonons and spins that allows us to disentangle the phonon and spin dynamics.  In cases where this cannot be done, we expect the phonon and spin dynamics to remain entangled. In other words, phonons have observable effects on any spin measurements which introduces spin operators away from the {\it Ising quantization axis} such as most entanglement witness operator measurements.
Finally, we may ask what does the entanglement evolution operator do to the system?  It is difficult to find any simple analytic estimates of the effect of this term, but it acts on the initial state which has the spins aligned along or opposite to the magnetic field and has no phonons.  During the evolution of that operator, new terms will be created which involve entanglement of spin states with states that have created phonons.  If the amplitude of those extra terms is small, they will not have a large effect, but if it is not, then one has no other recourse but to examine the full problem numerically, which is what we do next. First we examine a perturbation-theory treatment, and then we consider the full numerical evolution of the system.

\subsection{Diabatic effects from time-dependent spin-spin interaction}
One may have noticed that the spin evolution operator was not the evolution of a static Ising spin model.  There were additional time-dependent factors in the evolution operator which arose from the additional time dependence of the exchange operators that was inherited by the phonons when they were ``adiabatically'' removed from the problem. In this section, we use adiabatic perturbation theory (reviewed in Appendix B) to analyze the effect of those extra time-dependent terms on the spin evolution of the system.  In an adiabatic quantum simulation, one initially prepares the system in a certain pure state $|n(t_0)\rangle$ of the initial Hamiltonian $H(t_0)$ with the occupation $\alpha_{n}(t_{0})=1$ and the probability amplitudes in all other states vanishing [$\alpha_{m}(t_{0})=0$].
Thereafter, the probability amplitudes to be excited into the other states can be approximated by
\begin{equation}
\label{eq:excitations}
\alpha_{m}(t)\approx \int_{t_{0}}^{t}dt^{\prime} \frac{\langle m(t^{\prime})|\partial_{t^{\prime}}H(t^{\prime})|n(t^{\prime})\rangle}{E_{m}(t^{\prime})-E_{n}(t^{\prime})}e^{i[\theta_{m}(t^{\prime})-\theta_{n}(t^{\prime})]},
\end{equation}
for later times, as long as the transition amplitudes $|\alpha_{m}(t)|$ are much smaller than one during the time evolution.
This is the main expression we will use to evaluate the diabatic effects due to the time-dependent exchange interactions $J_{jj^{\prime}}(t)$.
Here $E_{m,n}(t')$ are the instantaneous eigenstates
of the spin Hamiltonian $H_{spin}(t)$ with a static exchange interaction $J_{j,j'}(t)=J_{j,j'}^{0}$ and $\theta_{m,n}(t')$ are the
corresponding dynamic phases given by the integrals $\theta_{m,n}(t)=\int_{t_{0}}^{t}dt^{\prime}\frac{E_{m,n}(t^{\prime})}{\hbar}$, and we assume there are no degeneracies in the instantaneous spectrum as a function of time.

Let us briefly describe the experimental protocol for a typical trapped ion quantum simulator restricted to the spin-only Hamiltonian $H_{spin}(t)=H_{Is}(t)+H_{B}(t)$ defined in Eq.~(\ref{eq:Hspin}).
The system is initially prepared in a spin-polarized state $|\uparrow_{y}...\uparrow_{y}...\uparrow_{y}\rangle$ along (or opposite to) the direction of the transverse magnetic field ${\bf B(t)}=B(t)\hat y$ by optical pumping followed by a $\pi/2$ spin rotation. The spin-only Hamiltonian is then turned on with a maximum effective transverse magnetic field $|{\bf B}|=B_m \gg |J_{jj'}(t)|$ followed by an exponential ramping down of the magnetic field to a final value $B_{m}e^{-t/\tau}$ at time $t$ ($\tau$ is the exponential ramping time constant for the decay of the magnetic field). After evolving to time $t$, the projection of the spin states along the $x$-axis of the Ising Hamiltonian is taken to find the probability to be in a particular spin state at time $t$ (in actual experiments another $\pi/2$ pulse is applied to rotate the $x$-axis to the $z$-axis where the measurement is made).

If the system is perfectly adiabatic during the evolution, the outcome of the quantum state would be the highest excited state
of the Ising Hamiltonian $H_{Is}(t)$ if the simulation starts out in the highest excited state of the magnetic field Hamiltonian  $B_{m}\sum_{j}\sigma_{j}^{y}$ at time $t=t_{0}=0$, which corresponds to the spins aligned along the $y$-axis. This procedure is theoretically identical to the ground state passage of the spin polarized state $|\downarrow_{y}...\downarrow_{y}...\downarrow_{y}\rangle$
to the ground state of the negative of the Ising Hamiltonian $-H_{Is}(t)$ with the system Hamiltonian being modified as $H(t)\rightarrow -H(t)$~\cite{Islam}.
In a typical trapped ion quantum simulator, the frequency $\mu$ is sufficiently far from any phonon frequencies such that the condition ${\eta_{X\nu}\Omega_{i}} \ll {|\mu-\omega_{X\nu}|}$ holds to avoid the heating of the system away from the initial phonon vacuum
state during the simulation. In addition, the maximum magnetic field strength is much larger than the time independent exchange interactions $|B_{m}|\gg |J^{0}_{jj'}|$ to ensure the system initially starts in an eigenstate of the initial Hamiltonian. To optimize the adiabaticity of the simulation, the ramping time constant $\tau$ for the magnetic field
has to be chosen to be much greater than the largest characteristic time scale of the system, which is shown below to be the minimum of the
inverse of the frequencies $|\mu-\omega_{X\nu}|, B_{m}/\hbar$.

We now discuss the effects of the time-dependent exchange interaction $\Delta J_{jj'}(t)$. For concreteness, we will follow the highest energy state, starting from the spin state aligned along the direction of the magnetic field.
Starting with the expression for the transition probability amplitude $\alpha_{m}(t)$ in Eq.~(\ref{eq:excitations}), we find the dominant diabatic transition
is to the state with the minimum energy difference $\Delta E=E_m(t_0)-E_n(t_0)$ with the initial spin polarized state $|\uparrow_{y}...\uparrow_{y}...\uparrow_{y}\rangle$, assuming the matrix element in the numerator does not depend too strongly on $\langle m|$, which is true when $H_{B}(t)\gg H_{Is}(t)$. At the initial time $t=0$, where the Ising couplings $J_{jj'}(t=0)$ can be shown to always vanish, all of the spin states with one spin flipped along the $y$-axis are
degenerate. This degeneracy will be broken by the Ising Hamiltonian $H_{Is}$ at finite time $t>t_0$. Due to spin-spin interaction in  $H_{Is}(t)$, the states along the y axis of the Bloch sphere called $|m\rangle$, have nonzero matrix components
$\langle m|\partial_{t^{\prime}}H_{Is}(t^{\prime})|n\rangle$
with the lowest energy gap $\Delta E \approx -2 B(t_0)$ with respect to the initial spin state $|n\rangle=|\uparrow_{y}...\uparrow_{y}...\uparrow_{y}\rangle$.

To approximately evaluate the transition amplitude $\alpha_{m}$ from the initial state to the two spin-flipped states $|m\rangle$, we do not actually need to know the state $|m\rangle$. The only relevant information we need is that it is one of the two spin-flipped states which tells us what the denominator is.
Hence, we can approximate
\begin{equation}
 \frac{\langle m(t^{\prime})|\partial_{t^{\prime}}H_{Is}(t^{\prime})|n(t^{\prime})\rangle}{E_{m}(t^{\prime})-E_{n}(t^{\prime})}
 \approx - \frac{\langle m|\partial_{t^{\prime}}H_{Is}(t^{\prime})|n\rangle}{2B(t)}.
\end{equation}
Using similar reasoning, we approximate $\Delta \theta(t^{\prime})$ as
\begin{equation}
\Delta\theta(t^{\prime})\approx - \int_{0}^{t^{\prime}}dt\frac{2B_{m}}{\hbar} e^{-\frac{t}{\tau}}=-2\omega_{B}\tau(1-e^{-\frac{t^{\prime}}{\tau}}),
\end{equation}
in which $\omega_{B}=\frac{B_{m}}{\hbar}$ is the magnetic angular frequency.
The operator $\partial_{t^{\prime}}H_{Is}(t^{\prime})=\sum_{j,j'}{\partial_{t^{\prime}} J_{j,j'}(t^{\prime})}\sigma_{j}^{x}\sigma_{j'}^{x}$ consists of modes with frequencies $\omega_{X\nu}+\mu$, $\omega_{X\nu}-\mu$, and $2\mu$
with the time derivative $\partial_{t^{\prime}} J_{j,j'}(t^{\prime})$ given by
\[
\partial_{t^{\prime}} J_{j,j'}(t^{\prime})=\hbar\sum_{\nu=1}^{N}
\frac{\Omega_{j}\Omega_{j'}\eta_{X\nu}^{2}b_{j}^{X\nu}b_{j'}^{X\nu}}
{\mu^2-\omega_{X\nu}^2}\mu\omega_{X\nu}
\]
\[
\times \Big[{\sin 2\mu t^{\prime}-{\cos \omega_{X\nu}t^{\prime}}{\sin\mu t^{\prime}} -\frac{\mu}{\omega_{X\nu}}\sin{\omega_{X\nu}t^{\prime}}
\cos{\mu t^{\prime}}}\Big].
\]
\begin{equation}
\approx \frac{\hbar}{2}\sum_{\nu=1}^{N}
\frac{\Omega_{j}\Omega_{j'}\eta_{X\nu}^{2}b_{j}^{X\nu}b_{j'}^{X\nu}}{\mu^2-\omega_{X\nu}^2}\mu\omega_{X\nu} (1-\frac{\mu}{\omega_{X\nu}})
\sin{(\omega_{X\nu}-\mu)t^{\prime}}.
\end{equation}
The last approximate expression is derived by keeping the contribution
from the slow mode $\omega_{X\nu}-\mu$ and dropping the
high frequency modes $\omega_{X\nu}+\mu$, and $2\mu$ because
the detuning $\mu>0$ is closely detuned to certain phonon modes in the quantum simulation.

As a consequence, the probability amplitude $\alpha^{Is}_{m,\nu}(t)$ induced by a single phonon mode
is given by the expression
\begin{equation}
\alpha_{m,\nu}^{Is}(t)=-\frac{i}{8}\sum_{jj'}\Big[ \frac{\mu\Omega_{j}\Omega_{j'}\eta_{X\nu}^{2}b_{j}^{X\nu}b_{j'}^{X\nu}}{\omega_{B}(\mu+\omega_{X\nu})}\Big]\langle
m|\sigma_{j}^{x}\sigma_{j'}^{x}|n\rangle f(t).
\end{equation}
The function $f(t)=2 i
\int_{0}^{t}dt^{\prime}
\sin{(\omega_{X\nu}-\mu)t^{\prime}}
e^{i\Delta\theta(t^{\prime})+\frac{t^{\prime}}{\tau}}$ can be approximated in experiments [when $\omega_{X\nu}-\mu$ is much larger than $\omega_{B}$ at slow ramping $(\omega_{X\nu}-\mu)\tau \gg 1$] as
\begin{equation}
f(t)\approx \frac{2}{i(\omega_{X\nu}-\mu)}
\left[e^{i\Delta\theta(t)+\frac{t}{\tau}}
\cos{(\omega_{X\nu}-\mu)t^{\prime}}
\Big|_{0}^{t}
\right].
\end{equation}
We therefore reach the conclusion that the probability amplitude $\alpha_{m,\nu}^{Is}(t)$ is given by
\[
\alpha_{m,\nu}^{Is}(t)=-\frac{1}{4}\sum_{j,j'}\langle m|\sigma_{j}^{x}\sigma_{j'}^{x}|n\rangle
\frac{\mu\Omega_{j}\Omega_{j'}\eta_{X\nu}^{2}b_{j}^{X\nu}b_{j'}^{X\nu}}{\omega_{B}(\omega_{X\nu}^2-\mu^2)}
e^{i\Delta\theta(t)+\frac{t}{\tau}}
\]
\begin{equation}
\times \left[1-\cos(\omega_{X\nu}-\mu)t\right].
\end{equation}
We note that diabatic effects manifested in $\alpha_{m,\nu}^{Is}$ due to time-dependent Ising couplings grow exponentially in time as $e^{t/\tau}$ signifying that the theory is only accurate for short times. To suppress the diabatic effects,
the criterion that has to hold for all phonon modes $\nu$ is
\begin{equation}
\left|\frac{1}{4}\sum_{\nu}\sum_{j,j'}\langle m|\sigma_{j}^{x}\sigma_{j'}^{x}|n\rangle
\frac{\mu\Omega_{j}\Omega_{j'}\eta_{X\nu}^{2}b_{j}^{X\nu}b_{j'}^{X\nu}}{\omega_{B}(\omega_{X\nu}^2-\mu^2)}e^{\frac{t}{\tau}}\right|\ll 1.
\end{equation}
Based on this expression, when the laser is closely detuned to one of the phonon resonance frequencies $\omega_{X\nu}$, the transition probability between states caused by $J_{j,j'}(t^{\prime})$ becomes large (diabatic). In addition, a stronger
magnetic field is required to suppress the diabatic transitions with smaller
detuning $\mu$. This is supported by the following numerical discussion in section {III \rm C}.
Notice that the above expression should be a reasonable estimation as long as the condition $B(t) \gg {\rm max}|J_{j,j'}(t)|=J_{max}$ applies
at time $t$ after the beginning of the quantum simulation. We can estimate the maximal time $t_{c}$ for which
$B(t) \gg {\rm max}|J_{j,j'}(t)|=J_{max}$ holds. The cutoff $t_c$ is set by $B(t_{c})=J_{max}$ where $J_{max}$ is the absolute value for the maximum exchange
interaction $J_{j,j'}(t)$ between the spins. As a result, the cut-off time $t_{c}$ is proportional to the ramping time constant $\tau$ with a logarithmic factor given approximately
by
\begin{equation}
t_{c}=\tau \ln \Big[\frac{B_{m}}{J_{max}}\Big].
\end{equation}
In the parameter regime where $B_{m}/J_{max} \gg 1$, in which our theory holds, $t_{c}$ can be extended somewhat beyond the ramping time constant $\tau$. Based on our numerical discussion, the diabatic effects
are largest when the magnetic field is ramped through the transition from paramagnetic state to other targeted spin states, which is also accompanied by larger phonon creations due to the shrinkage of the spin gap near the transition.

\section{Numerical results}
In this section, we focus on showing the circumstances where quantum emulators can or cannot be described by the transverse-field Ising model with high fidelity. Since our goal is to understand under what circumstances the effect of the phonons is small, we consider different cases for the time-evolution of the system including various detunings and initial transverse magnetic field strengths.

To isolate different effects, we compare two spin-only models in the presence of the ramping magnetic fields with the theoretically exact spin-phonon model based on numerical diagonization.
The first is the {\it ideal spin model} which considers the evolution of the system with a static Ising model (spin-exchange coefficients are the time-averaged exchange coefficients) and a time-dependent magnetic field. While one might think this is a purely adiabatic model, it has some diabatic effects, since the fully polarized state is {\it not} generically the ground state of the Ising plus magnetic field Hamiltonian because the (static)  Ising exchange interactions are nonzero at the initial time. Hence, one can invoke a sudden approximation to the system initially, and find that the initial state is a superposition of different energy eigenstates.  In addition, the magnetic field varies in time and hence can cause additional diabatic effects due to its derivative with respect to time.

The second is the {\it effective spin model} which involves, essentially, evolution of the spin system according to the spin evolution operator only in Eq~(\ref{eq:Hspin}).  Hence it has the static Ising Hamiltonian, the time-dependent Ising interactions and the time-varying transverse magnetic field.  This model can have its Schr\"odinger equation solved in a spin-basis only, since all phonon effects are neglected except virtual
phonon excitations.

The third model is the {\it exact spin-phonon model}, where we evolve the system according to the original spin-phonon Hamiltonian expanded in the Lamb-Dicke limit [Eq. (\ref{eq:phonon})]. The only approximation used in this last model is the cutoff for the phonons.
The strategy we use is to numerically integrate the Schr\"odinger equation using a direct product basis which involves a spin state in direct product with a phonon state. We do this because the Hamiltonian only connects states that differ by plus or minus one phonon number, and hence is block sparse in this basis. The spin states are chosen to include all possible Ising spin states for the number of ions in the trap.  The phonon basis is chosen to have a cutoff of a maximal phonon excitation. The maximal cutoff is always chosen to be larger than the average occupancy of the phonons in each normal mode of the ion chain.  Of course, we expect more phonons to be excited into the phonon modes closest to the beatnote frequency of the lasers, so the cutoffs that are chosen will vary from one normal mode to another.  For example, we often find we can set the phonon cutoff to be one for some of the phonon modes far from the driving frequency of the spin-dependent force.

To facilitate our discussion, we define
the root-mean square average $J_{rms}$
 of the fully connected Ising interaction for $N$ ions as
\begin{equation}
J_{rms}=\sqrt{\frac{\sum_{j,j'}|J_{j\neq j'}|^{2}}{N(N-1)}},
\end{equation}
in which the static Ising interaction $J_{j\ne j'}$ is given by the static term in Eq.~(\ref{eq:exchange}) and the integer indices $j,j'$ both range from $1$ to $N$.

\subsection{Symmetry and experimental protocols}
Let us discuss the symmetry of the spin-only system first, which is relevant for the exact diagonization of the spin-only Hamiltonian.
There is one spatial inversion symmetry (${\bf R}_j\rightarrow -{\bf R}_j$) in the ion chain, since the equilibrium ion positions are distributed symmetrically about the origin in the trap and all phonon modes involve symmetric or antisymmetric displacements of corresponding ion positions.
There is also a spin reflection symmetry ($\sigma^{x}\rightarrow -\sigma^{x}$, $\sigma^y\rightarrow \sigma^y$, and $\sigma^z\rightarrow -\sigma^z$) in the spin-only models with a transverse magnetic field $B(t)\sum_{j}\sigma_{j}^{y}$.  This spin-reflection symmetry preserves all commutation relations of the spin operators and leaves the Hamiltonian invariant. Therefore, there are four symmetry sectors for the eigenstates of the spin model (even space, even spin; even space, odd spin; odd space, even spin; and odd space, odd spin).

 If the static Ising couplings are all negative (positive), the spin ground state is ferromagnetic (antiferromagnetic)
and the highest spin eigenstate is the opposite, namely antiferromagnetic (ferromagnetic). We will focus on a detuning to the blue of the center-of-mass phonon mode.  In this case, all spin exchange couplings are positive and the ground state is antiferromagnetic, while the highest excited state is ferromagnetic. We will examine the adiabatic state evolution of the highest eigenstate.
With all the respected discrete symmetries, we can construct the symmetric
and antisymmetric ferromagnetic states of the spin-only Hamiltonian as $\frac{1}{\sqrt 2}(|\uparrow_{x} \cdots \uparrow_{x}\rangle+|\downarrow_{x}\cdots\downarrow_{x}\rangle)$ or
$\frac{1}{\sqrt 2}(|\uparrow_{x} \cdots \uparrow_{x}\rangle-|\downarrow_{x}\cdots\downarrow_{x}\rangle)$ which is in the (even, even) or (even, odd) sectors, respectively.

The experimental protocol is to prepare the system initially in a spin polarized state $|\uparrow_{y}\cdots\uparrow_{y}\rangle$,
[which is the highest eigenstate of the transverse magnetic field $B(t)\hat y$], by optical pumping and a coherent spin rotation and then to gradually turn off the magnetic field with an exponential ramp $B(t)=B_{m} e^{-t/\tau}$ while keeping the spin-dependent laser force in the $x$-direction on
 during simulation time through stimulated Raman transitions between the spin states. According to adiabatic evolution, if the quantum state is initially prepared in the highest eigenstate of the field-only Hamiltonian $B(t)\sum_{j}\sigma_{j}^{y}$, the outcome of the quantum simulation will adiabatically follow the corresponding highest eigenstate of the Hamiltonian (Ising spin Hamiltonian), if there are no level crossings (which does not occur in this system).
 In the case with positive static Ising coupling, the ferromagnetic highest energy eigenstate is the
 symmetrical ferromagnetic entangled state (the so-called GHZ state) $\frac{1}{\sqrt 2}(|\uparrow_{x} \cdots \uparrow_{x}\rangle+|\downarrow_{x}\cdots\downarrow_{x}\rangle)$, when $B>0$.

There are two intrinsic errors which  can impede the quantum simulation in trapped ions.
The first is diabatic effects which occur primarily when either parameters in the Hamiltonian are changed too rapidly in time, or when energy gaps in the instantaneous eigenvalue spectrum become to small.
The second is the error induced by phonons in the presence of time-dependent transverse magnetic fields. For example, the phonon-spin Hamiltonian does not have spin-reflection symmetry because it is linear in the $\sigma^x$ operators, and hence the spin-phonon interaction breaks this $Z_{2}$ symmetry. One consequence of this is to couple the symmetric and antisymmetric ferromagnetic states which is likely to reduce the spin entanglement of the GHZ state.
(Other errors such as phonon decoherence effects due to spontaneous emission are not considered here.)

\subsection{Probability and GHZ state entanglement measurements}

Current experiments use atomic cycling transitions to measure the spin state of the ion (which clock state the ion is in), and do not measure the phonons excited in the system.  Hence,
the experimental observables are the probability $P_{\beta}(t)$ of a spin-polarized state after tracing out phonons in the tensor product of the spin-phonon Hilbert space $|spin\rangle \otimes|phonon\rangle,$ as mentioned above when discussing Eq.~({\ref{eq:prob}}). If one performs rotations about the Bloch sphere prior to making the measurement of the probabilities, then one can also measure a number of different spin-entanglement witness operators.

A spin-entanglement witness operator (for a target entanglement state)~\cite{Witness} is a mathematically constructed observable that has a negative expectation
value when the system is entangled. No witness operator can measure general entanglement, but instead a witness operator is constructed to measure a specific type of spin entanglement.
For example, the witness operator $W_{{\rm GHZ}}$ for an $N$-ion chain can be constructed as~\cite{Witness}
\begin{equation}
\label{eq:witness}
W_{{\rm GHZ}}=3I-2\Big[\frac{S_{1}^{{\rm GHZ_{N}}}+I}{2}
+\prod_{k=2}^{N}\frac{S_{k}^{{\rm GHZ_{N}}}+I}{2}\Big]
\end{equation}
with the stabilizing spin operators expressed in terms of the Pauli spin operators by
\begin{equation}
S_{1}^{{\rm GHZ_{N}}}=\prod_{k=1}^{N}\sigma^{y}_{k},\quad
S_{k}^{{\rm GHZ_{N}}}=\sigma^{x}_{k-1}\sigma^{x}_{k}.
\end{equation}
Notice that the target spin polarized state in this paper is along the Ising $x$ axis in the Bloch sphere instead of the $z$ axis. Therefore, we modified the original expression~\cite{Witness} to our problem by the transformation $\sigma^{z}\rightarrow\sigma^{x}$ and $\sigma^{x}\rightarrow\sigma^{y}$.
Based on the above construction, GHZ state entanglement measurements
can be detected by the observable $\langle W_{{\rm GHZ}}\rangle=Tr(\rho W_{{\rm GHZ}})$ with the density matrix $\rho$ constructed by pure states or mixed states during
the quantum simulation. For a perfect GHZ state entanglement, one can show that the entanglement witness operator satisfies  $\langle W_{{\rm GHZ}}\rangle=-1$  (refer to Appendix C). Any deviation from perfect GHZ entanglement would lead to a value greater than $-1$.  Note that this is one of the few cases of a witness operator where the degree of entanglement is correlated with the magnitude of the expectation value of the witness operator.

\subsection{Transition to the ferromagnetic state when detuned blue of the center-of-mass phonon mode}

The systems we consider range from $N=4$ to $N=8$ which is far from the thermodynamic limit. The quantum phase transition (QPT) due to the discontinuity of the ground-state wave function in the thermodynamic limit $N\rightarrow \infty$ only manifests itself as a state avoiding crossing in the energy spectrum, which is adiabatically connected to the QPT at large $N$. The system parameters and the higher set of transverse phonon modes, which belongs to the higher branch of two transverse motional degrees of freedom, for different numbers of ions $N$
are summarized in Table I. The trapping parameters are given by the aspect ratio and the CM mode frequency $\omega_{CM}$ along the transverse (tight) axis. The axial (easy) trapping frequency is given by the product of the aspect ratio and the CM mode frequency $\omega_{CM}$. The choice of these parameters comes from trap parameters and typical operating regimes of the ion-trap experiment at the University of Maryland.  Most results are robust with moderate changes of parameters and our choices do not intimate that fine tuning of parameters is needed to achieve the results we show.

\begin{table}[htbp!]
\caption{Parameter set I} % title name of the table
\centering % centering table
\begin{tabular}{|c|c c c|} % creating 10 columns
\hline% inserting double-line
Aspect ratio & &0.2092  &
\\
\hline % inserts single-line
$\omega_{CM}/2\pi$ & & $4.63975$~MHZ &   \\
\hline
Rabi frequency $\Omega/2\pi$ & & $369.7$~KHZ &   \\
\hline
Lamb Dicke parameter $\eta_{CM}$ & & 0.06  &   \\
\hline
No. of ions N & 4 & 6 & 8  \\
\hline
 & 1.0 & 1.0 & 1.0     \\
& 0.9788 \ & 0.9788 \ & 0.9788    \\
Transverse&0.9482 \ & 0.9481 \ & 0.9480  \\
phonon mode  &  0.9088 \ & 0.9083 \ & 0.9079  \\
frequencies & &0.8589 \ & 0.8581  \\
(in units of $\omega_{CM}$) & &0.7988 \ & 0.7974  \\
& & &  0.7236 \\
& & &  0.6324 \\
\hline % inserts single-line
\end{tabular}
\label{tab:Parameters1}
\end{table}

\begin{figure}[htbp!]
\centering
\includegraphics[width=3.5in,height=4.0in]{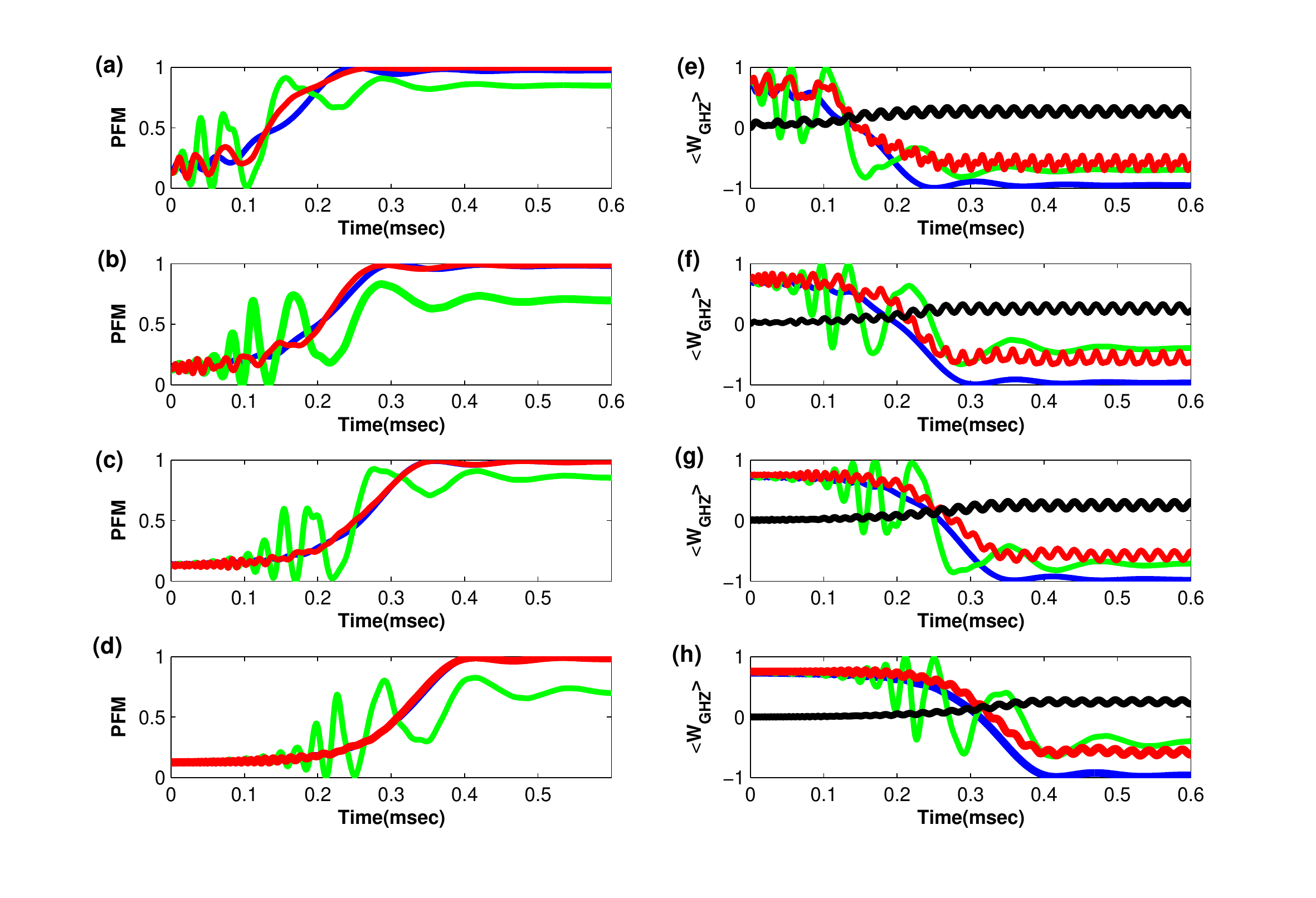}\\
\caption{(Color online.) Quantum simulation for different models (blue line: {\it ideal spin model}, green line: {\it effective spin model}, red line: {\it exact spin-phonon model}). The average phonon occupation number $\bar n(t)$ is plotted with a solid black line in the right panels (e--h). The right panels also show the expectation value of the GHZ entanglement witness operator.  The detuning $\mu$ is always chosen to be
$1.0095\omega_{CM}$, just slightly blue of the CM phonon mode, while the exponential ramping time constant for the magnetic field is fixed at $\tau = 8\times 10^{-2}$~msec. The initial transverse magnetic field varies in the different panels as follows:
(a) $B_{m} = 23.64J_{rms}$,
(b) $B_{m} = 47.28J_{rms}$,
(c) $B_{m} = 94.56J_{rms}$, and
(d) $B_{m} = 189.12J_{rms}$.
The average static Ising coupling (in units of angular frequency)
is given by  $J_{rms}=1.5017\times 10^{-4}\omega_{CM}$ (for the chosen ion trap parameters in Table I).
}
\label{fig:FIG1}
\end{figure}

\begin{figure}[htbp!]
\centering
\includegraphics[width=3.5in,height=4.0in]{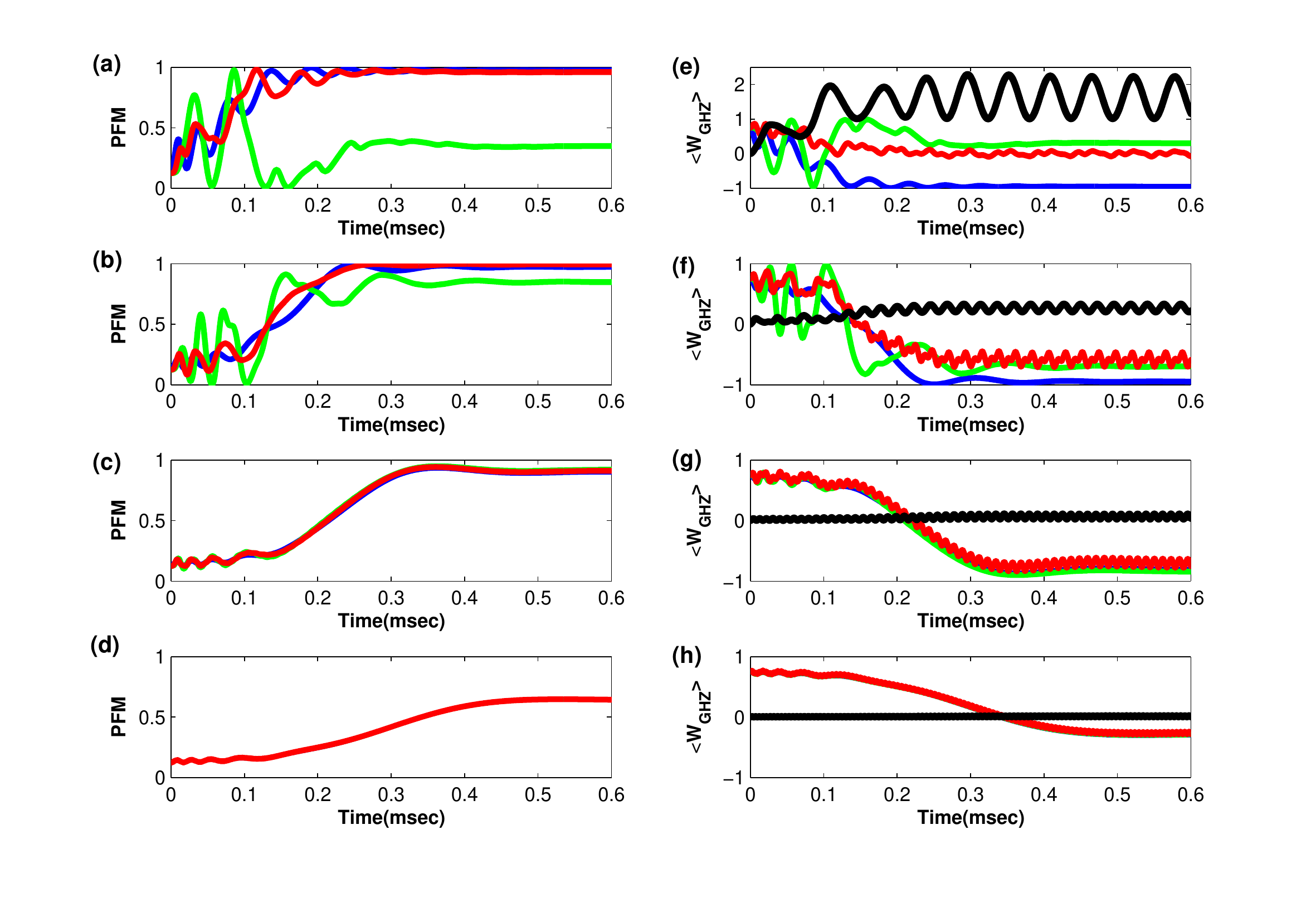}\\
  \caption{ (Color online.) Detuning dependence $\mu$ of the simulation (the initial magnetic field is $B_{m}= 3.6\times 10^{-3}\omega_{CM}$ and the exponential ramping time constant $\tau=8\times 10^{-2} msec$ are both fixed) for
 $\mu=1.0038\omega_{CM}, J_{rms}=3.765\times 10^{-4}\omega_{CM}$~(a),
 $\mu=1.0095\omega_{CM}, J_{rms}=1.5017\times 10^{-4}\omega_{CM}$~(b),
 $\mu=1.019\omega_{CM}, J_{rms}=7.4734\times 10^{-5}\omega_{CM}$~(c),  and $\mu=1.038\omega_{CM}, J_{rms}=3.7019\times 10^{-5}\omega_{CM}$~(d).
The other panels (e--h) are the corresponding GHZ witness operator and the average phonon number.}
  \label{fig:FIG2}
\end{figure}

In Fig.~\ref{fig:FIG1}, the left panel shows the probability to be all spins up, or all spins down  [PFM$(t) = |P_{\uparrow_{x}
 \uparrow_{x} \cdots \uparrow_{x}}(t)|^2+|P_{\downarrow_{x}\downarrow_{x} \cdots \downarrow_{x}}(t)|^2$] and the right panel shows the corresponding entanglement witness operator expectation value $\langle W_{GHZ}\rangle$ for the GHZ state with four ions ($N=4$), positive (blue) detuning ($\mu>\omega_{CM}$) and positive Ising coupling ($J>0$).
 In this case, the highest excited state is ferromagnetic and the center of mass (CM) phonon mode produces nearly uniform Ising couplings between each spin (more uniform the closer the detuning is to the CM frequency).

We begin by examining the effect of the magnitude of the initial magnetic field on the adiabatic state evolution.  In theory, we want the magnetic field as large as possible, because the larger it is, the better the initial state is an eigenstate of the system.  But this limit must be balanced against two competing issues: first, we must complete the simulation in a finite time period and we do not want parameters in the Hamiltonian to change too rapidly (to prevent large diabatic effects) and second, the experimental setup can only create a magnetic field of some finite maximum value.  Hence, we must always live with a finite initial magnetic field.  If the field is too low, then the sudden approximation tells us we will have larger admixtures of excited states in the system, which will give rise to oscillations in the measured probabilities, and will reduce the final probability to end up in the ferromagnetic state. This is because the initial state is not close to the initial eigenstate of the system.  If the field magnitude is much less than the spin coupling, then the system never evolves to the ground state, but remains in the state ordered along the magnetic field.

The set of curves in panel (a) of Fig.~\ref{fig:FIG1} is at an intermediate value of the magnetic field. Let us focus first on the {\it ideal spin model}.  One can see that the probability has initial oscillations, which continue as the system evolves, but it produces a nearly pure ferromagnetic state probability by the end of the simulation.  The corresponding witness operator shows that we achieve excellent entanglement as well.  If we add the time-dependent exchange interactions as described by the {\it effective spin model}, we see the probabilities generate much larger diabatic oscillations, and the final probability is strongly suppressed.  In addition, the witness operator gives an expectation value shifted significantly up from $-1$. Finally, when we consider the exact coupled spin-phonon model, we find the oscillations lie in between those of the ideal and {\it effective spin models}, but the probability for the ferromagnetic state is nearly one.  The witness operator, on the other hand, shows reduced entanglement, similar to the {\it effective spin model}. This is
easier to understand for the spin-phonon model, because the presence of phonons breaks the $Z_2$ symmetry of the spin reflection, and hence the classification of states as even or odd is lost, so those states are mixed together by the Hamiltonian reducing the entanglement.  Furthermore, comparing to the phonon occupancy, we see deviations between the red and green curve start when the phonon occupation is still quite low, and the oscillations in the witness operator are strongly correlated with oscillations in the phonon number (because the phonon and spin are coupled in a correlated fashion, of course).  As the field is increased for fixed ramping time (panels b through d), we see that the amplitude of the oscillations is reduced for the {\it ideal spin model} and for the {\it exact spin-phonon model}, but the witness operator for the exact solution always shows reduced entanglement. This behavior provides one of our main conclusions about the effect of the phonons---namely, they act to provide a decoherence of the spin system, which reduces oscillations and improves the adiabatic nature of the evolution of the probabilities, but reduces the overall entanglement of the spin state.

\begin{figure}[htbp!]
\centering
\includegraphics[width=3.5in,height=4.0in]{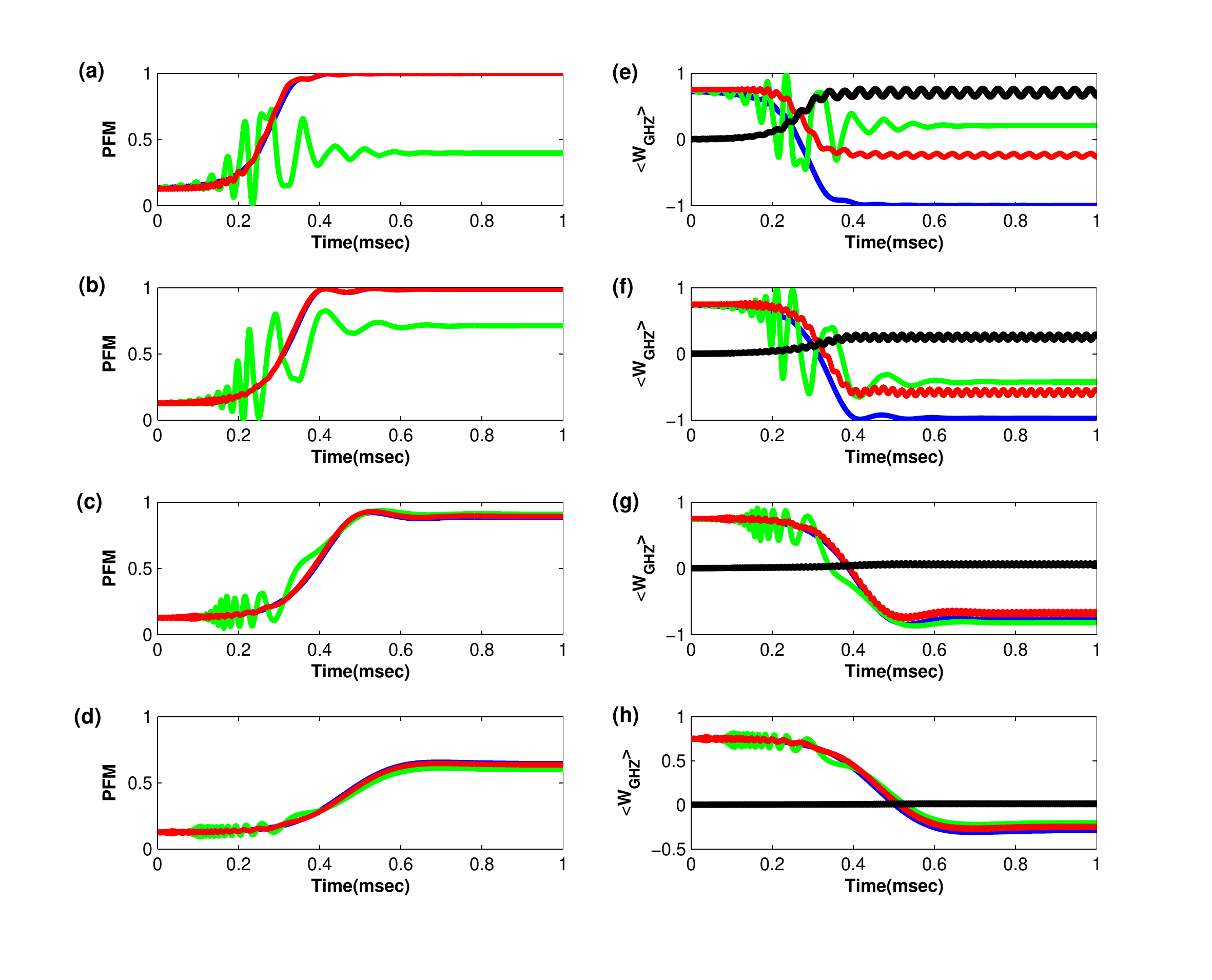}\\
  \caption{(Color online.) Detuning dependence of the simulation with a strong initial magnetic field
($B_{m}= 2.88\times 10^{-2}\omega_{CM}$ and $\tau=8\times 10^{-2}$~msec) for
$\mu=1.0057\omega_{CM}, J_{rms}=2.5076\times 10^{-4}\omega_{CM}$~(a),
$\mu=1.0095\omega_{CM}, J_{rms}=1.5017\times 10^{-4}\omega_{CM}$~(b),
$\mu=1.0190\omega_{CM}, J_{rms}=7.4734\times 10^{-5}\omega_{CM}$~(c), and $\mu=1.0380\omega_{CM}, J_{rms}=3.7019\times 10^{-5}\omega_{CM}$~(d). Panels (e--h) are the corresponding GHZ witness operator and the average phonon number.
}
  \label{fig:FIG3}
\end{figure}

In Fig.~\ref{fig:FIG2}, we examine the detuning dependence of the simulator in the case of a moderate magnetic field.  We start with the detuning very close to the CM mode, and then move further and further away in panels (a) through (d). As expected, when we are far detuned from the phonon line (panel d), no phonons are created, and all of the three models agree essentially perfectly for the probability and the entanglement.   As we move in closer to the phonon line, we start to generate phonons and we also generate diabatic oscillations in the probability data, but surprisingly, we continue to see a very close correspondence of the {\it exact spin-phonon model} to the {\it ideal spin model}, while the {\it effective spin model} describes the system in an increasingly poorer fashion.  The entanglement gets sharply reduced, as well, so much so that for the smallest detuning, the entanglement witness operator is effectively zero for the {\it exact spin-phonon model}. As before the {\it ideal spin model} is incorrect in predicting the entanglement, and the {\it effective spin model} is much better at approximating the entanglement.

\begin{figure}[htbp!]
\centering
\includegraphics[scale=0.4]{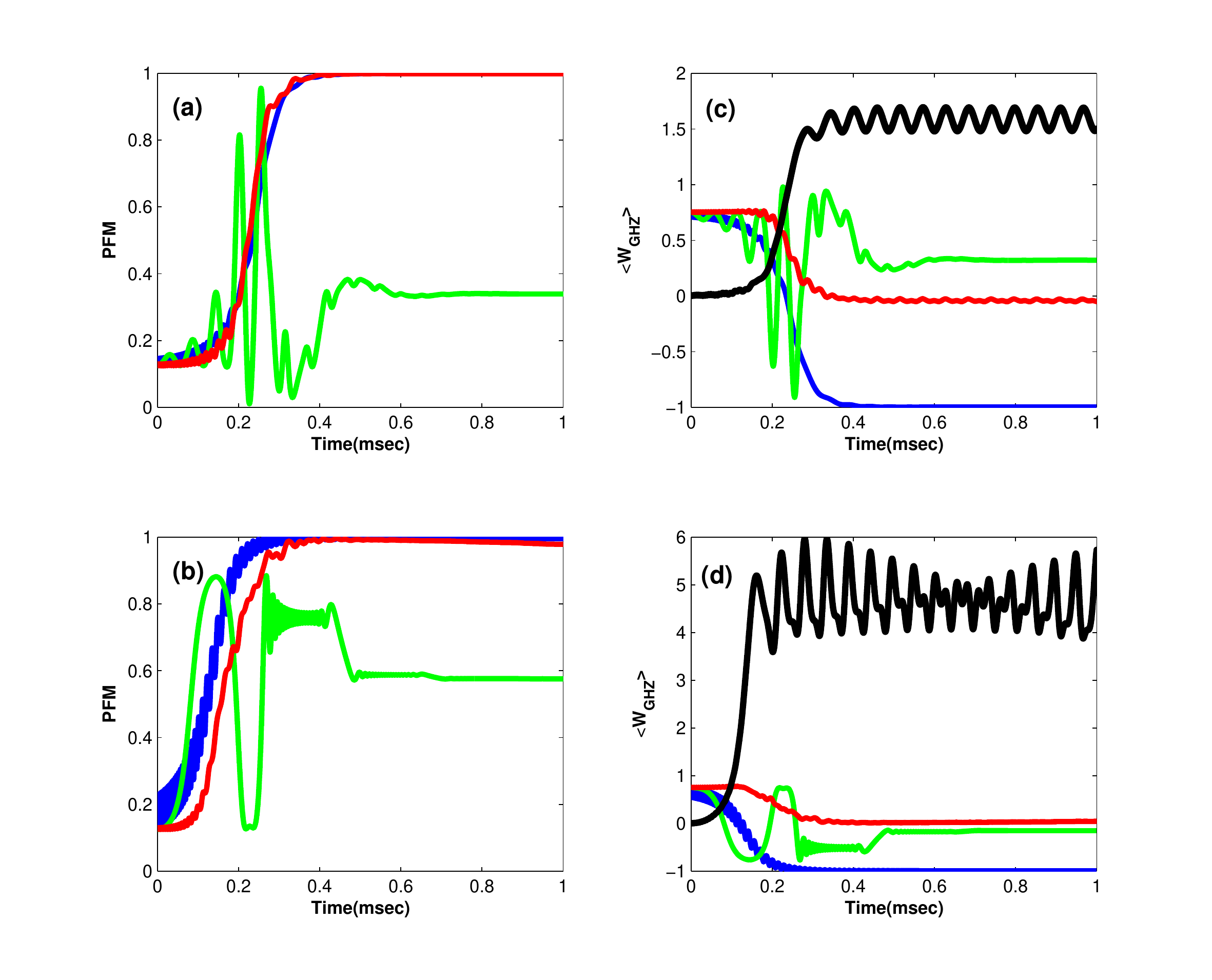}\\
  \caption{(Color online.) Quantum simulation with detuning near resonance with the CM mode
and with a large initial magnetic field
($ B_{m}= 2.88\times 10^{-2}\omega_{CM}$ and $\tau=8\times 10^{-2}$~msec) for
$\mu=1.0038\omega_{CM}, J_{rms}=3.765\times 10^{-4}\omega_{CM}$~(a) and $\mu=1.00095\omega_{CM}, J_{rms}=1.5\times 10^{-3}\omega_{CM}$~(b). Panels (c--d) are the corresponding GHZ witness operator and the average phonon number.
}
  \label{fig:FIG4}
\end{figure}

In Fig.~\ref{fig:FIG3}, we also examine the detuning dependence, but now in the case of a large magnetic field. Here the {\it exact spin-phonon model} and the {\it ideal spin model} give virtually identical results for the ferromagnetic probability and the {\it effective spin model} fails once the detuning gets too close to the phonon line. The entanglement, however, continues to behave as before, with the {\it ideal spin model} always predicting large entanglement and the {\it exact spin-phonon model} showing much reduced entanglement.  Note that the trend of reducing the ferromagnetic probability and the entanglement as the detuning moves away from the phonon line is arising from the fact that the exchange parameters are getting smaller, so the field is evolving the system too rapidly to be in the adiabatic limit and we have diabatic effects which do not allow us to reach the ferromagnetic state during the time of the simulation.

\begin{figure}[htbp!]
\centering
\includegraphics[width=3.5in,height=4.0in]{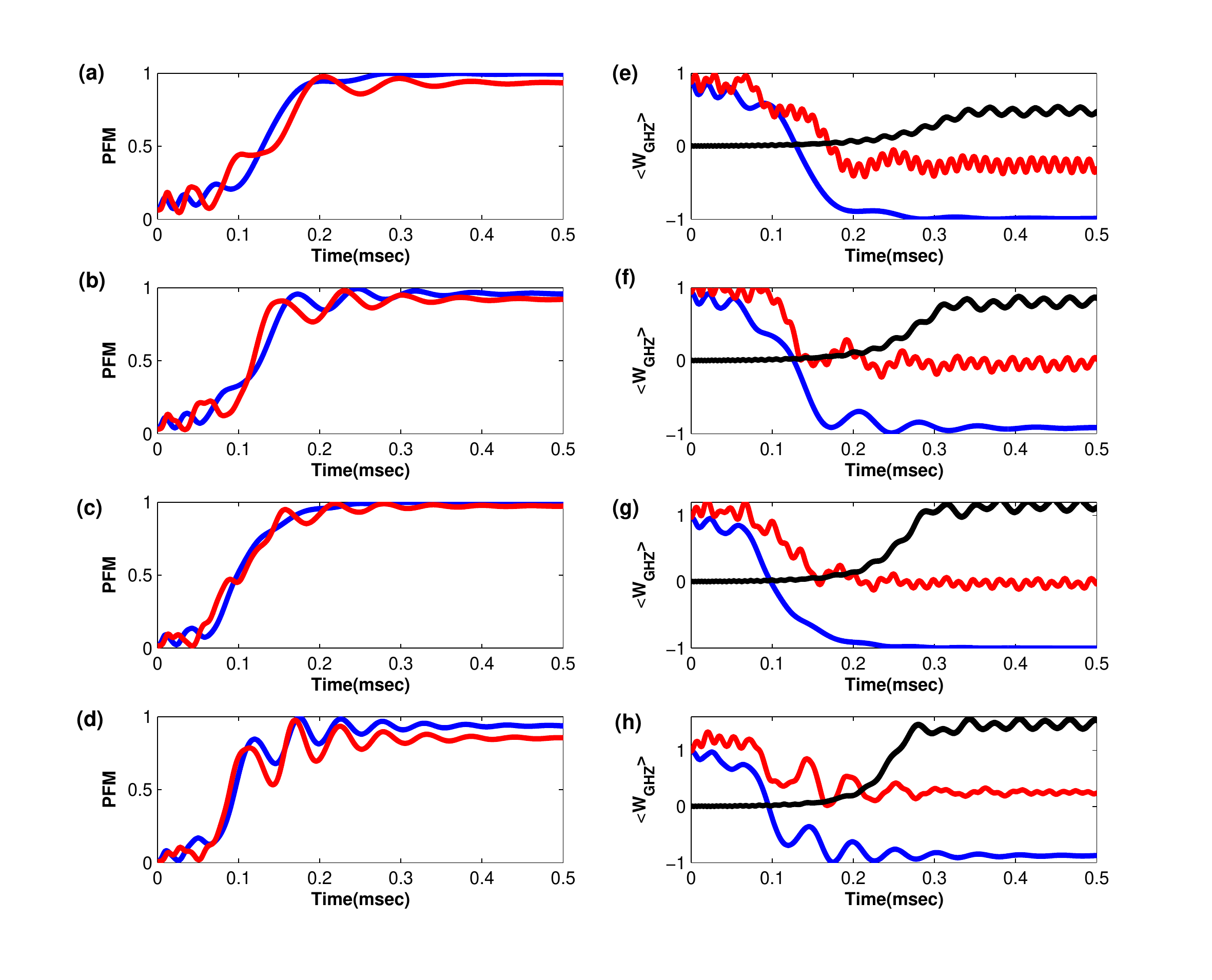}
  \caption{(Color online.) Number dependence of quantum simulation for intermediate initial magnetic field and detuning chosen according to the text.
  In the left panel subplots, we plot the ferromagnetic probability as a function of time. In the right panel subplots, we plot the time dependence of the GHZ state witness operator $\langle W_{\rm GHZ}\rangle$. Red solid lines are results for the {\it exact spin-phonon model} and blue solid lines are for {\it ideal spin model}. The black lines indicate  the corresponding coherent CM phonon occupation driven by the lasers.
   The initial magnetic field strength $B_{m}=3.6\times 10^{-3}\omega_{CM}$ and exponential ramping time constant  $\tau=8\times 10^{-2}$~msec are chosen with varying parameters given by the following cases:
   (a) $N=5,\mu=1.0076\omega_{CM}, J_{rms}=1.5013\times 10^{-4}\omega_{CM}$,
   (b) $N=6, \mu=1.0063\omega_{CM}, J_{rms}=1.5011\times 10^{-4}\omega_{CM}$,
   (c) $N=7, \mu=1.0054\omega_{CM}, J_{rms}=1.5006\times 10^{-4}\omega_{CM}$, and
   (d) $N=8, \mu=1.0048\omega_{CM}, J_{rms}=1.5008
  \times 10^{-4}\omega_{CM}$. Panels (e--h) are the corresponding GHZ witness operator and the average phonon number.}
  \label{fig:FIG5}

\end{figure}

If we go to the case of near resonant driving of the system, as in Fig.~\ref{fig:FIG4}, we see that we create significant numbers of phonons, and we eventually see that the {\it exact spin-phonon model} does deviate from the {\it ideal spin model} in the probabilities.  As expected, we also see a very sharp reduction of the entanglement of the exact evolution as compared to the {\it ideal spin model}.

So far, all of our calculations were for a small chain with $N=4$.  Now we look at the size dependence of the ion chain by examining the behavior as we increase the number $N$ from 5 to 8 in Fig.~\ref{fig:FIG5}. This case is with a moderate initial magnetic field, but driving fairly close to resonance with the CM mode.  In order to properly compare the different systems, we adjust the detuning to yield an average spin exchange that is approximately equal for the different number of ions. These results clearly show the main themes we have been exploring.  The {\it exact spin-phonon model} and the {\it ideal spin model} exhibit nearly identical evolution of the probability, with the deviations starting to become clear only for the largest chain.  Nevertheless, the entanglement of the system is sharply reduced for all systems, with the reduction occurring at about the same time in the simulation for all cases. In terms of CM mode phonon excitations, we observe
the trend of increasing phonon excitations from far below one phonon toward more than one phonon as the system size increases. The strong correlation between phonon excitations and the deviation between the {\it exact spin-phonon model} and the {\it ideal spin model} is still
clearly manifested.

\begin{figure}[htbp!]
\centering
\includegraphics[width=3.5in,height=4.0in]{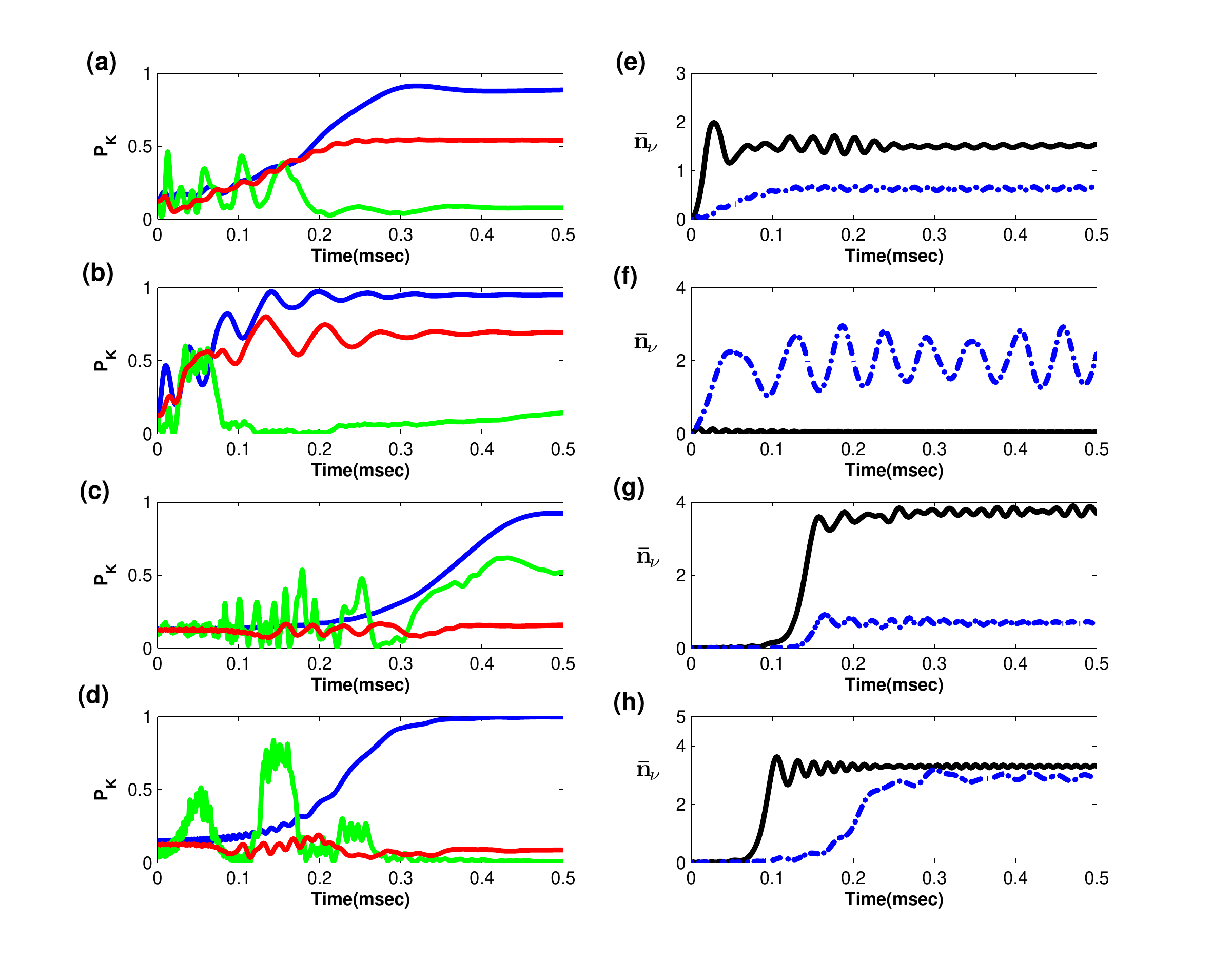}
  \caption{(Color online.) Quantum simulation for the  kink state in the presence of  phonon excitations. The left panel shows the probability of the one-kink state $P_{K}$ as a function of time for the  {\it ideal spin model} (blue), {\it effective spin model} (green) and the {\it exact spin-phonon model} (red) for different cases. The right panels show the average phonon occupation number $\bar n_{\nu}(t)$ for the CM mode (black lines) with angular frequency $\omega_{CM}=2\pi\times4.6398$~MHZ and the tilt mode (blue lines) with angular frequency $\omega_{T}=0.9788\omega_{CM}$.
  We fix the exponential ramping time constant $\tau=8\times 10^{-2}$~msec for different magnetic field and detunning as:
  (a) $\mu=0.9905\omega_{CM}$, $B_{m}=3.6\times 10^{-3}\omega_{CM}$,
  (b) $\mu=0.9810\omega_{CM}$, $B_{m}=3.6\times 10^{-3}\omega_{CM}$,
  (c) $\mu=0.9905\omega_{CM}$,  $B_{m}=2.88\times 10^{-2}\omega_{CM}$, and (d) $\mu=0.9810\omega_{CM}$, $B_{m}=2.88\times 10^{-2}\omega_{CM}$.
Panels (e--h) are the corresponding GHZ witness operator and the average phonon number. (System parameters are chosen according to Table I.)
}
  \label{fig:FIG6}
\end{figure}

\subsection{ Evolution to the kink phase in the highest excited state}
Let us look at the highest excited state on ion chains with even numbers of ions and a detuning that lies between the highest (CM) mode and the second (tilt) mode.  We examine this scenario here to see the effects of phonons on this transition.

In Fig.~\ref{fig:FIG6}, we show predictions for experimental traces when the detuning is between these two phonon lines. The red detuning is chosen such that it is possible to excite two transverse phonon modes simultaneously by locating the laser detuning between the CM mode  on an ion chain with $N=4$ ions $\omega_{CM}\approx 2\pi \times 4.64$ MHz and the tilt mode with frequency $\omega_{T}= 0.9788\omega_{CM}$.
The spin state simulation ideally should evolve from the highest excited state $|\uparrow_{y}...\uparrow_{y}...\uparrow_{y}\rangle$ along the transverse magnetic field axis
to the symmetric kink state (or antiphase state) $\frac{1}{\sqrt 2}(|\uparrow_{x}\uparrow_{x}\downarrow_{x}\downarrow_{x}\rangle
+|\downarrow_{x}\downarrow_{x}\uparrow_{x}\uparrow_{x}\rangle)$ along the Ising axis if the system is perfectly adiabatic.
In those cases,
long range spin-spin couplings
are negative and dominant over adjacent spin couplings; therefore,
the highest excited state is the symmetric kink state (with ferromagnetic couplings, the highest excited state will be antiferromagnetic).
By detuning from the CM mode $\omega_{CM}$ to the tilt mode $\omega_{T}$, the adjacent spin-spin couplings change sign from negative to positive and long range spin-spin couplings are still negative with enlarged magnitude. We have checked that this eigenstate is generic for even number of ions (up to at least $N=8$) with the highest eigenstate always being the antiphase (kink) state.

In cases (a) and (b), with the same magnetic field ramping, we observe a
larger probability for the kink phase $P_{K}$ in panel (b) (further detuned from the CM mode) for the {\it ideal spin model}.
Similarly, we also observe a larger $P_{K}$ (higher fidelity) in case (d) than in case (c).
Lower fidelity in general is correlated to the fact that the spin excitation gap is smaller (with respect to the exponential ramping time constant $\tau$). This implies that the frustration of the Ising couplings favors a larger spin gap for intermode detuning  than that for the close detuning case to the blue of the CM mode. By exact numerical diagonization for {\it ideal spin models},
we found it is indeed the case and the second highest eigenstate
has very different character. In cases (a) and (c), the second
highest eigenstate is the symmetric antiferromagnetic state
$\frac{1}{\sqrt 2}(|\uparrow_{x}\downarrow_{x}\uparrow_{x}\downarrow_{x}\rangle
+|\downarrow_{x}\uparrow_{x}\downarrow_{x}\uparrow_{x}\rangle)$
at the end of the simulation
with a spin gap $\Delta_{s}\approx 2\pi\times 28.359$KHZ .
However, for cases (b) and (d),
the second highest eigenstate is the kink state
$\frac{1}{2}(|\uparrow_{x}\downarrow_{x}\downarrow_{x}\downarrow_{x}\rangle
+|\downarrow_{x}\uparrow_{x}\uparrow_{x}\uparrow_{x}\rangle+
|\downarrow_{x}\downarrow_{x}\downarrow_{x}\uparrow_{x}\rangle+
|\uparrow_{x}\uparrow_{x}\uparrow_{x}\downarrow_{x}\rangle
)$ with a larger spin gap $\Delta_{s}\approx 2\pi\times 107.40$KHZ.

As far as phonon effects are concerned, we notice that the large
 discrepancies between the {\it ideal spin model} and the {\it exact spin-phonon} calculations for $P_K$ are strongly correlated to the large number of phonon excitations (${\bar n}_{\nu}>1$) for either phonon mode.
By comparing case (a) with case (c) [or comparing case (b) with case (d)], we notice that large phonon excitation often occurs during the part of the evolution when the
 magnetic field strength is larger.
Unlike the  blue detuned situation in subsection C,  a small
detuning from the CM mode does not show high fidelity of the final spin state (kink order in this case) at long times. Instead, the probability for observing the kink state is significantly degraded by the presence of the phonons.  Hence, the creation of phonons does not always help the system appear as if it is simulating a static spin Hamiltonian.

By greatly reducing the Rabi frequency  to suppress the phonon excitations, we observe good agreements between {\it ideal spin models} and
{\it exact spin phonon models} in case (a) as well as the enhancement of the spin probability $P_{K}$. For larger ion number case ($N=6$), we find
the diabatic effects completely ruin the simulation by tuning accessible
system parameters in Table I to an order of magnitude larger or smaller.
Even if one can suppress the phonon effects in this case, we will still typically have large diabatic effects due to the small energy gaps.

\subsection{Kink transition in the ground state}
\begin{figure}[htbp!]
\centering
\includegraphics[scale=0.4]{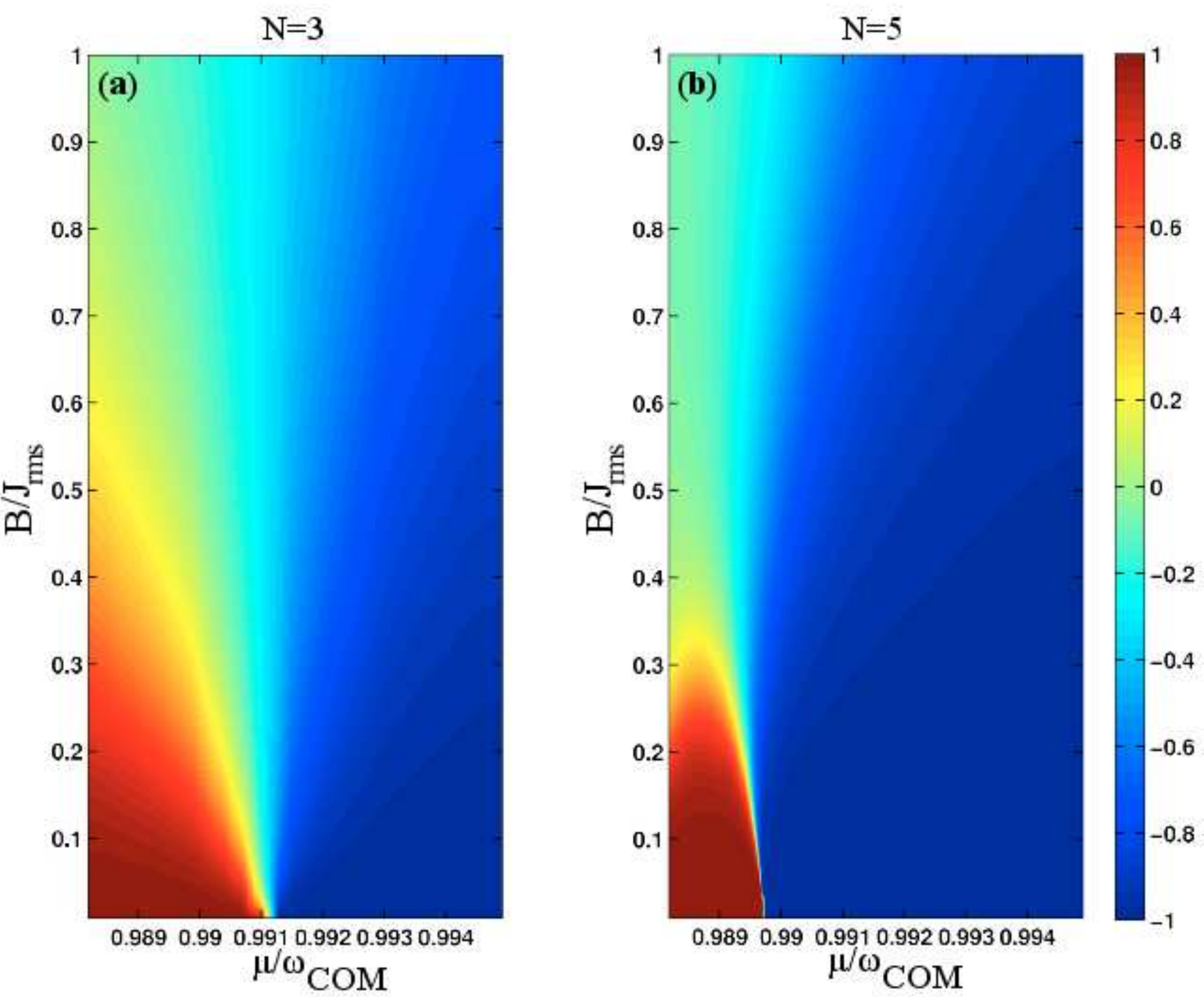}\\
\caption{(Color online.)
  Adiabatic phase diagram of the transverse field Ising model for  (a) $N=3$ (left panel) and (b) $N=5$ (right panel).
  The horizontal axis is the laser detuning scaled by the transverse CM  mode
  frequency $\omega_{CM}$. The vertical axis is the transverse magnetic field $B$ scaled by the root-mean-square average of Ising coupling $J_{rms}$.
  The blue area represents the  one-kink phase and the red area indicates the ferromagnetic phase. The range of the detuning $\mu$ (in units of  $\omega_{CM}$) is shown between the second phonon mode and the third phonon mode. The value of the order parameter $P_{F}-P_{K}$
 ( varying from $-1$ to $+1$) is described by the color scale to the right of the figure.
}
  \label{fig:FIG7}
\end{figure}

Following the ground state evolution with an odd number of ions and a laser detuning between the second and third highest transverse phonon modes, there is a sharp phase transition~\cite{LM}
separating ferromagnetic and one-kink spin ground states for the {\it ideal spin models}.
Therefore, it is quite relevant to examine the feasibility for experimentally observing this transition by going beyond an adiabatic evolution  of the system and treating phonon effects and diabatic effects exactly.
The crossover becomes sharp very quickly with the number of ions due to
the fully-connected nature of the spin models.
A typical adiabatic phase diagram for the spin only Hamiltonian with $N=3$ and $N=5$ is shown in Fig.~\ref{fig:FIG7}. The order parameter $P_{F}-P_{K}$ for the transition is defined as the difference between the probabilities of the two degenerate ferromagnetic states $P_{F}$ and the four degenerate one-kink states $P_{K}$ .
The one-kink states for the $N=3$ case are given by the four degenerate states
$|\uparrow_{x}\downarrow_{x}\downarrow_{x}\rangle$, $|\downarrow_{x}\uparrow_{x}\uparrow_{x}\rangle$,
$|\downarrow_{x}\downarrow_{x}\uparrow_{x}\rangle$, and
$|\uparrow_{x}\uparrow_{x}\downarrow_{x}\rangle$.
For $N=5$, the four one-kink states
 $|\uparrow_{x}\uparrow_{x}\downarrow_{x}\downarrow_{x}\downarrow_{x}\rangle$
,$|\downarrow_{x}\downarrow_{x}\uparrow_{x}\uparrow_{x}\uparrow_{x}\rangle$,
$|\downarrow_{x}\downarrow_{x}\downarrow_{x}\uparrow_{x}\uparrow_{x}\rangle$, and
$|\uparrow_{x}\uparrow_{x}\uparrow_{x}\downarrow_{x}\downarrow_{x}\rangle$
are also degenerate.
It is clear  that the crossover between the two phases is
sharper (smaller spin gap) for $N=5$ than $N=3$ as shown by the much sharper contrast in color.
In addition, the ferromagnetic phase area moves closer to the leftmost phonon mode and shrinks in size as the number of ions increases.
As the magnetic field gets smaller, the transition between the ferromagnetic and kink phases becomes very sharp as $N$ increases~\cite{LM}. The numerical discussion in this section is based
on the system parameters in Table II.

\begin{table}[htbp!]
\caption{Parameter set II} % title name of the table
\centering % centering table
\begin{tabular}{|c|cc|} % creating 10 columns
\hline% inserting double-line
Aspect ratio & \ \ \ \  0.1 &
\\
\hline % inserts single-line
$\omega_{CM}/2\pi$ &\ \ \ \  $4.63975$~MHZ  &   \\
\hline
Rabi frequency $\Omega/2\pi$ & \ \ \ \ $3.697$~KHZ &   \\
\hline
Lamb Dicke parameter $\eta_{CM}$ & \ \ \ \ 0.06  &   \\
\hline
No. of ions N &\hspace*{-10mm} 3 &\hspace*{-10mm} 5  \\
\hline
Transverse & \hspace*{-10mm} 1.0 & \hspace*{-10mm} 1.0     \\
phonon mode frequencies & \hspace*{-10mm} 0.9950  & \hspace*{-10mm} 0.9950     \\
(in units of $\omega_{CM}$) &\hspace*{-10mm} 0.9879 & \hspace*{-10mm} 0.9879 \\
 &\hspace*{-10mm}   &\hspace*{-10mm} 0.9789  \\
 & \hspace*{-10mm} & \hspace*{-10mm} 0.9683\\
\hline % inserts single-line
\end{tabular}
\label{tab:Parameters2}
\end{table}

\begin{figure}[htbp!]
\centering
\includegraphics[scale=0.6]{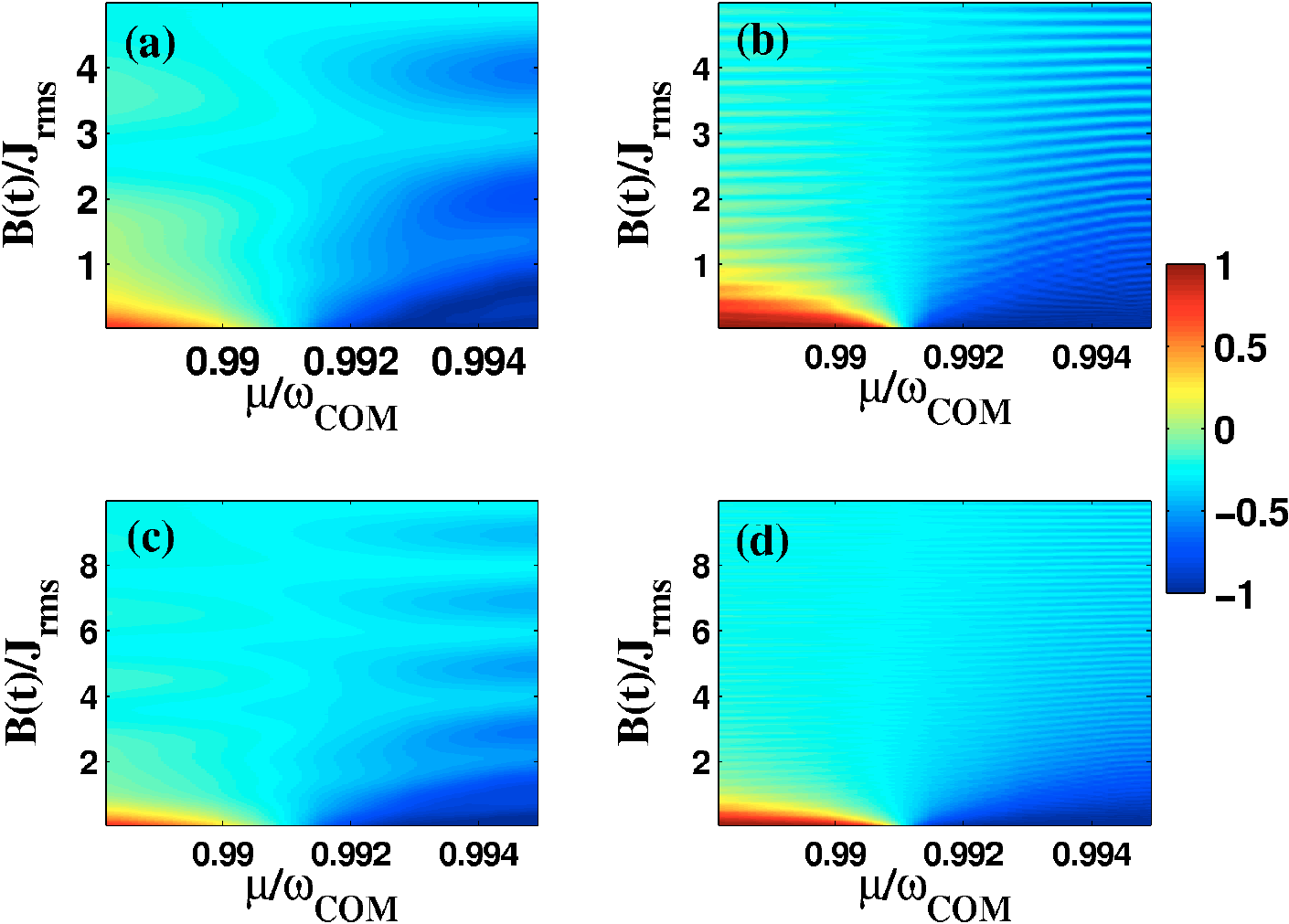}\\
\caption{(Color online.) Ferromagnetic to kink phase diagram for $N=3$ calculated for the ideal spin model with diabatic effects included.
The horizontal axis is the laser detuning $\mu$ scaled by transverse CM frequency $\omega_{CM}=2\pi\times4.6398$MHZ.
The vertical axis is the instantaneous transverse magnetic field $B(t)$ scaled by the root-mean-square average of the spin couplings $J_{rms}$ (note the range changes for different panels).
The Rabi angular frequency $\Omega$ and the dimensionless Lamb-Dicke parameter for the center of mass mode $\eta_{CM}$ are selected to be $\Omega=0.01\omega_{CM}$ and $\eta_{CM}=0.06$ respectively.
  (a) $\tau=0.2T, T=0.614\times(J_{rms}/2\pi)^{-1}, B_{m}=5J_{rms}$,
  (b) $\tau=0.2T, T=6.14\times(J_{rms}/2\pi)^{-1}, \tau=0.2T, B_{m}=5J_{rms}$,
  (c) $\tau=0.2T, T=0.614\times(J_{rms}/2\pi)^{-1}, \tau=0.2T, B_{m}=10J_{rms}$, and
  (d): $\tau=0.2T, T=6.14\times(J_{rms}/2\pi)^{-1}, B_{m}=10J_{rms}$.               }
  \label{fig:FIG8}
\end{figure}

In Fig.~\ref{fig:FIG8}, we numerically map out the time dependence of the probability $P_{F}-P_{K}$ for {\it ideal spin models}.
The nexponentially ramped magnetic field $B(t)$ is chosen with different initial values $B_{m}$  (scaled by $J_{rms}$ as determined by the detuning $\mu$ and the Rabi angular frequency $\Omega$).
The value of the Rabi angular frequency $\Omega=0.01\omega_{CM}$ is chosen so that it is safely within the weak field regime $\eta_{\nu}\Omega_{\nu}\ll (|\omega_{X\nu}-\mu|)$ near the central region of the phase diagram for all phonon modes. The total simulation time
$T$ is chosen so that it is proportional to the inverse of $J_{rms}/2\pi$.
We select the exponential ramping time constant $\tau$ for the exponential reduction of the
magnetic field $B(t)$ to be one-fifth of the experimental simulation time
($\tau=0.2 T$). By comparing (a) to (b) [or (c) to (d)], we observe that diabatic effects are greatly suppressed when the exponential ramping time constant $\tau$ is large enough so that the transition to the closest excited state in energy is negligibly small. This effect shows up as  much deeper colors dictating the order parameter $P_{F}-P_{K}$ in the ferromagnetic states and the kink states when $B(t)$ approaches zero on the vertical
axis, as illustrated in subplots (b) and (d). The diabatic effects also show up clearly as a slow oscillation in the probabilities $P_{F}-P_{K}$
at larger $B(t)/J_{rms}>1$  before the simulation ends along the vertical axis in subplots (a) and (c).
 We also notice some fast background oscillations in  $P_{F}-P_{K}$ covering the entire phase diagram in cases (c) and (d). This effect
is due to the fact that the time derivative of the dynamic phase $\theta_{m}(t)-\theta_{n}(t)$ between $(m,n)$ states is roughly stationary in time (as analyzed with adiabatic perturbation theory). For short exponential ramping time constants $\tau$, one cannot see the noticeable interference pattern
between these states because of a random phase cancelation along the path of the state evolution in time. In panels (b) and (d) the main difference is a reduction of the period of the background interference pattern, which is shorter in panel (b) (larger magnetic field).

\begin{figure}[htbp!]
\centering
\includegraphics[scale=0.6]{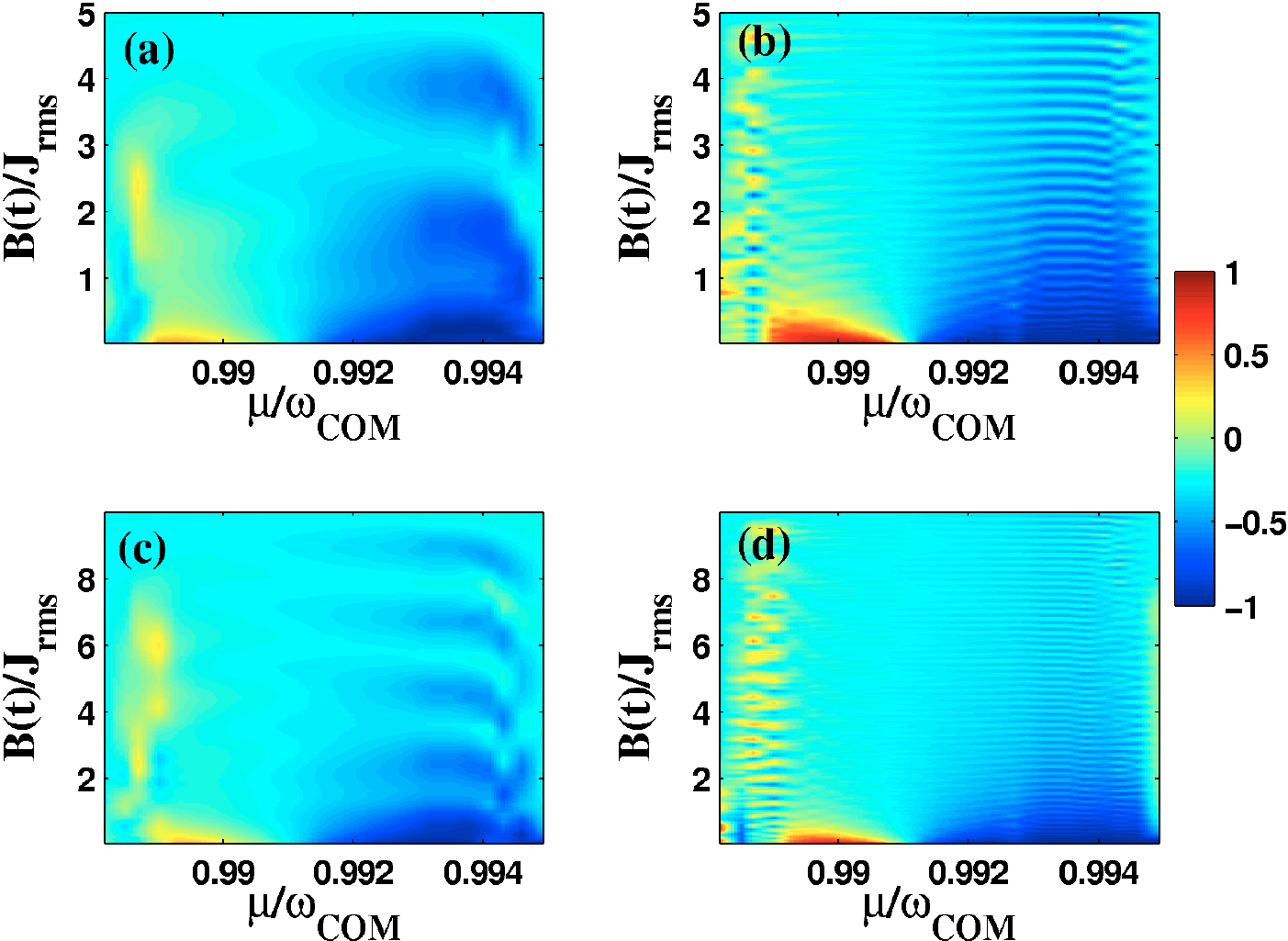}\\
\caption{(Color online.)
Ferromagnetic to kink phase diagram for $N=3$ with the same cases as in Fig~\ref{fig:FIG8}. The phase diagrams are calculated with the exact spin-phonon Hamiltonian.
The horizontal axis is the laser detuning $\mu$ scaled by transverse CM mode
frequency $\omega_{CM}$.
The vertical axis is the instantaneous transverse magnetic field $B(t)$ scaled by the root-mean-square average of spin-spin coupling $J_{rms}$.  (a): $\tau=0.2T, T=0.614\times(J_{rms}/2\pi)^{-1}, B_{m}=5J_{rms}$,
(b): $\tau=0.2T, T=6.14\times(J_{rms}/2\pi)^{-1}, \tau=0.2T, B_{m}=5J_{rms}$,
(c): $\tau=0.2T, T=0.614\times(J_{rms}/2\pi)^{-1}, \tau=0.2T, B_{m}=10J_{rms}$, and
(d): $\tau=0.2T, T=6.14\times(J_{rms}/2\pi)^{-1}, B_{m}=10J_{rms}$.             }
  \label{fig:FIG9}
\end{figure}

What are the phonon effects on the corresponding ferromagnetic to kink phase diagram? We show our calculations for the $N=3$ case in Fig.~\ref{fig:FIG9}. Phonon creation has serious effects on the  phase diagram. In the cases with fast ramping time constants  [(a) and (c)], the ferromagnetic states are destabilized and appear only with small probability.
For slow ramping time constants [cases (b) and (d)], the FM domain disappears near the leftmost phonon mode
due to large phonon creation  as the phonon is being more resonantly driven. But the kink state domain reduces only slightly near the rightmost phonon mode.
As a consequence, phonons restrict the available parameter space to observe the FM to kink phase diagram  but do not
rule out the possibility of observing the phase as long as
the exponential ramping time constant $\tau$ is long enough.
In the current numerical simulation we show, the exponetial  ramping time constant $\tau$ is roughly on the order of  a few milliseconds (close to feasibility in current experiments).
One may suspect that phonons can ruin the stability of the FM
state when the number of ions scales up
because the FM domain shrinks in size and moves closer to the leftmost phonon mode, and if we are too close to the phonon mode, phonon creation ruins the chance to see the FM state. One can try to increase the experimental simulation time and reduce the Rabi frequency, but doing this too much eventually runs into coherence issues or problems from spontaneous emission.
%In addition, by comparing subplots (b) and (d), we notice
%that phonons are more populated in the large onset magnetic field cases
%near the second phonon mode frequency and cause the FM state boundary get pushed toward the blue %side of the detuning $\mu$.

\begin{figure}[htbp!]
\centering
\includegraphics[scale=0.6]{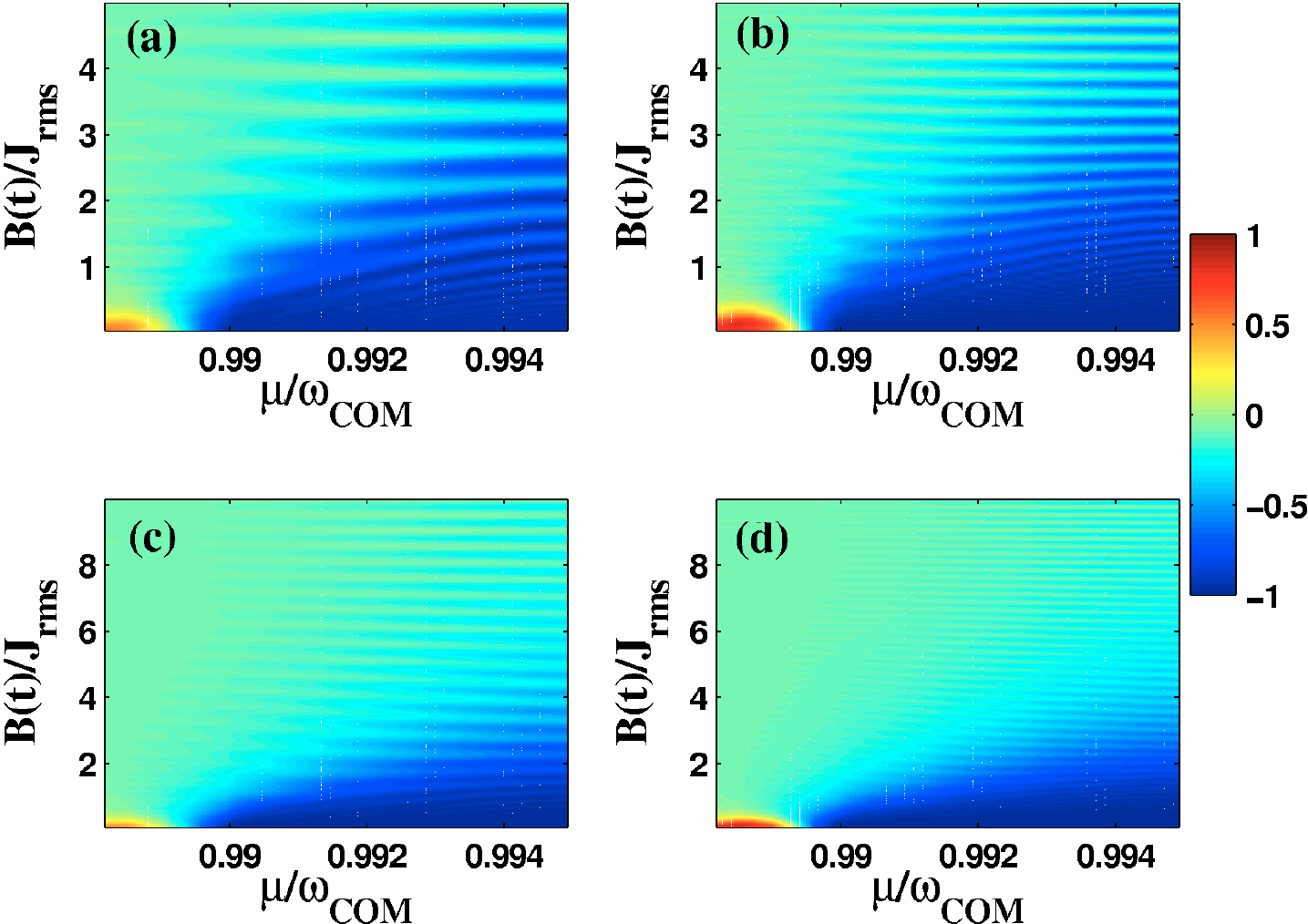}\\
\caption{(Color online.) Ferromagnetic-kink phase diagram for $N=5$ as calculated for the {\it ideal spin model}.
The horizontal axis is the laser detuning $\mu$ scaled by transverse mode trapping frequency $\omega_{CM}=2\pi\times4.6398$MHZ.
The vertical axis is the instantaneous transverse magnetic field $B(t)$ scaled by the root-mean-square average of spin-spin coupling $J_{rms}$.
The Rabi angular frequency $\Omega$ and the dimensionless Lamb-Dicke parameter for the CM mode $\eta_{CM}$ are $\Omega=0.0086\omega_{CM}$ and $\eta_{CM}=0.06$, respectively.
  (a) $\tau=0.2T, T=0.614\times(J_{rms}/2\pi)^{-1}, B_{m}=5J_{rms}$,
  (b) $\tau=0.2T, T=6.14\times(J_{rms}/2\pi)^{-1}, \tau=0.2T, B_{m}=5J_{rms}$, (c) $\tau=0.2T, T=0.614\times(J_{rms}/2\pi)^{-1}, \tau=0.2T, B_{m}=10J_{rms}$, and
  (d) $\tau=0.2T, T=6.14\times(J_{rms}/2\pi)^{-1}, B_{m}=10J_{rms}$.               }
  \label{fig:FIG10}
\end{figure}

In Fig.~\ref{fig:FIG10}, we show the $N=5$ case for the {\it ideal spin model}.
The behavior of the FM-kink phase diagram is similar to the $N=3$ case
in Fig.~\ref{fig:FIG8} except
the boundary of FM states and kink states is shifted toward lower
detunings $\mu$.
As a consequence, the FM state domain (deep red area) occupies
the region where the detuning is close to the leftmost phonon mode and the kink state domain (deep blue area) occupies most of the detuning region.
However, the phase diagram for low magnetic field $B(t)\rightarrow 0$ is very close to the adiabatic phase diagram (see the right panel of Fig.~\ref{fig:FIG7}) when the initial magnetic field $B_{m}$ is large and the ramping time constant $\tau$ is long as shown in Fig.~\ref{fig:FIG10} (b) and Fig.~\ref{fig:FIG10} (d)] except for the background interference patterns that were described above.
One also notices that there is much less diabatic effects at low $B(t)$ due to the fact that the smallest spin excitation gap is larger near the central area of the phase diagram.

\begin{figure}[htbp!]
\centering
\includegraphics[scale=0.6]{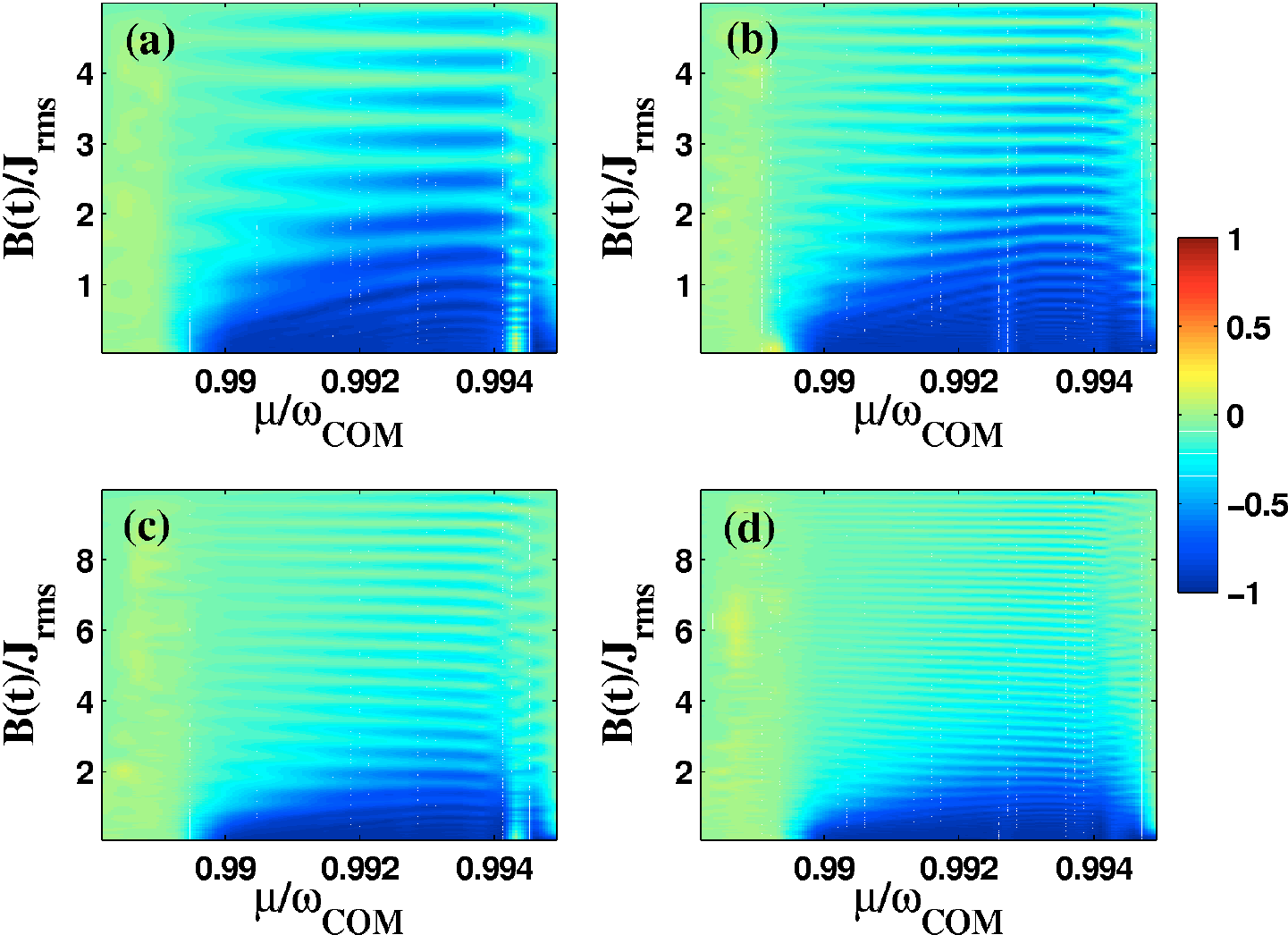}\\
\caption{(Color online.)
Exact FM-kink phase diagram for the spin-phonon Hamiltonian with $N=5$ and the corresponding parameter set as in  Fig.~\ref{fig:FIG10}.
The horizontal axis is the laser detuning $\mu$ scaled by the transverse CM  frequency $\omega_{CM}$.
The vertical axis is the instantaneous transverse magnetic field $B(t)$ scaled by the root-mean-square average of spin-spin coupling $J_{rms}$.  (a) $\tau=0.2T, T=0.614\times(J_{rms}/2\pi)^{-1}, B_{m}=5J_{rms}$,
(b) $\tau=0.2T, T=6.14\times(J_{rms}/2\pi)^{-1}, \tau=0.2T, B_{m}=5J_{rms}$, (c) $\tau=0.2T, T=0.614\times(J_{rms}/2\pi)^{-1}, \tau=0.2T, B_{m}=10J_{rms}$, and
(d) $\tau=0.2T, T=6.14\times(J_{rms}/2\pi)^{-1}, B_{m}=10J_{rms}$.             }
\label{fig:FIG11}
\end{figure}

When we add phonon effects, we might expect the phase diagram to only deviate when we are detuned close to a phonon line, but the situation is much worse for $N=5$, as shown in  Fig.~\ref{fig:FIG11}.
The kink phase (deep blue zone) exists for a wide range
of ramping and onset magnetic fields $B_{m}$ as shown
in all cases.
However,  the ferromagnetic domain (red zone) disappears even for slow ramping  [like the exponential ramping time constant $\tau\approx 30$~msec in cases (c) and (d)].
This does not rule out the possibility of observing the FM phase for
even longer ramping time constants $\tau$ (or smaller Rabi angular frequency $\Omega$) but a ramping time constant $\tau\approx 30$ msec is already well beyond what is used in current ion-trap experiments where $\tau$ is usually less than one millisecond.
This problem gets worse for larger $N$, and already for $N=7$ the FM-kink phase diagram appears to be impossible to observe.
This arises in part due to the fact that the spin gap closes exponentially fast with the system size~\cite{Duan}. As a result, one needs to dramatically reduce the diabatic effects to see the transition. In addition, phonon effects also make it hard to see the transition by not allowing the detuning to move too close to either phonon line, and thereby misses significant regions where the FM phase is stable.

\section{Discussion and conclusions}
We have examined a number of different issues related to the importance of phonons in analog quantum simulation of the transverse field Ising model.  We show when the spin-phonon entanglement operator $U_{en}$ can be approximated by $1$ (a longitudinal magnetic field along the Ising axis, or a vanishing transverse magnetic field), one can show that the phonons do not affect the probability to measure the spins in product states in the direction of the Ising interaction, but they can reduce the entanglement of the spin eigenstates.  Surprisingly, in cases when the operator $U_{en}$ cannot be approximated by $1$, the effect of the phonons is often to make the system look more like a static Ising spin Hamiltonian plus a time-varying transverse magnetic field.  This result holds primarily when laser detuning  is blue of the CM mode, and hence corresponds to a ferromagnetic case when one looks at the highest excited state.
We emphasize that the common belief based on the geometric phase gate, in which  phonon effects can be suppressed by choosing the period to be the inverse of close detuning from a phonon mode due to periodic spin-phonon entanglement dynamics, is no longer valid in a finite decaying transverse magnetic field.

Our work shows that one must consider phonon effects in most ion-trap spin simulator experiments especially when the spin-spin interaction is highly frustrated.  In cases when laser is detuned blue of the transverse center of mass mode, phonons are beneficial to make the system look more and more like the static spin Hamiltonian being emulated (at the expense of reducing spin-spin entanglement).  In cases when spin interactions are frustrated with multiple phonon modes stimulated, the phonons can work to suppress the true adiabatic spin phases from having a high fidelity or even invalidate the spin phases completely. Generically the phonon effects beyond adiabatic elimination are remarkable when the detuning lies close to at least one of the phonon frequencies, and hence on average more than one phonon per mode is excited.

In conclusion, large laser detuning is essential to suppress phonon coherent population while it also causes the shrinkage of spin excitation gaps in the adiabatic spin simulation. Alternative adiabatic quantum simulation schemes which do not create noticeable phonon occupation, while maintaining large spin excitation gaps would be desirable.

\section*{ACKNOWLEDGEMENTS}
We acknowledge interesting discussions with Kihwan Kim, Rajibul Islam, Wes Campbell, Emily Edwards, Simcha Korenblit, Zhe-Xuan Gong, Chris Monroe, and L.-M. Duan. Joseph Wang thanks Marcos Rigol for sharing computational resources.
This work was supported under ARO grant number W911NF0710576 with funds from the DARPA OLE Program. J.~K.~F.~acknowledges support of the McDevitt bequest at Georgetown.

\appendix

\section{Spin-dependent force and effective transverse magnetic field}
We describe how different laser-ion interactions are employed to ultimately simulate
effective spin models.
 The spin-dependent dipole force along any direction of the equatorial plane of the Bloch sphere and an effective transverse magnetic field are created  by using multiple laser beams and the optical-dipole interaction between the ions and phase-locked lasers.  We start with a
 reference Raman beam that has frequency $\omega_{L}$ and then superpose another perpendicular Raman beam (that has frequencies in a frequency comb $\omega_{L}+\omega_{0}+\mu, \omega_{L}+\omega_{0}-\mu, \omega_{L}+\omega_{0}$).  The lasers use off-resonant Raman coupling through dipole-allowed excited states of the ion to generate an effective spin-phonon interaction. Take the Ytterbium ion ${}^{171}\textrm{Yb}^{+}$  as an example: the qubit states are the clock states
 $|F=0,m_{F}=0\rangle$ and $|F=1,m_{F}=0\rangle$, formed from the hyperfine states of the $\textrm{S}$ valence electron and the spin one-half nucleus. These states have no linear Zeeman effect, and hence are less prone to background magnetic fields.

 The hyperfine state $|F=1,m_{F}=0\rangle$ is denoted as the spin-up state and the state  $|F=0,m_{F}=0\rangle$ is the spin-down state in the z-axis of the pseudospin Bloch sphere. The energy-level spacing is in the gigahertz range, so one could, in principle, directly make transitions that flip the spins from up to down and vice-versa by stimulated emission and absorption processes acting on the hyperfine states. But, it is common to instead generate these spin-flip transitions via off-resonant Raman coupling to a third state to suppress incoherent spontaneous emission effects. To do this we need two laser beams with different frequencies which can be detuned away from the energy level spacing between the clock states and each of the frequencies are chosen to be far away from dipole allowed resonant transitions in each ion.
We denote the two beams wavevectors and frequencies  by $({\bf k}_{1},\omega_{1}),({\bf k}_{2},\omega_{2})$ respectively. By adiabatic elimination~\cite{PJLee} of dipole-allowed excited states through the Raman procedure for ion $j$, one can write down the interaction for an ion as
\begin{equation}
H_{LI}=\frac{\hbar\Omega_{j}^{e}}{2}(S_{j}^{+}+S_{j}^{-})
\left[e^{i(\Delta {\bf k}\cdot{\bf {\hat R}}_{j}-\Delta\omega t+\phi)}+H.C. \right],
\end{equation}
where $\Omega_{j}^{e}$ is the effective Rabi frequency of the stimulated-Raman transition, $\hbar\Delta {\bf k}=\hbar({\bf k}_{1}-{\bf k}_{2})$ and $\hbar\Delta {\bf \omega}=\hbar(\omega_{1}-\omega_{2})>0$ are the effective momentum and energy of the photons, respectively,  $\phi$ is the controlled phase shift between the two laser beams,  and the pseudospin flip operators  are $S_{j}^{+},S_{j}^{-}$.

The full Hamiltonian involves the sum of this term plus the  clock state energy level difference $\hbar\omega_{0}$ multiplied by the $z$-component spin operator.
Now we go to the
interaction picture with respect to the clock state energy level difference $\hbar\omega_{0}$,
\begin{equation}
H_{LI}^{I}=\exp[\frac{i\omega_{0}}{2}\sigma_{j}^{z}t]H_{LI}
\exp[-\frac{i\omega_{0}}{2}\sigma_{j}^{z}t]
\end{equation}
at time $t$.
With the photon energy difference $\hbar\Delta\omega$ comparable to the clock-state energy splitting $\hbar\omega_{0}$, only terms with slow modes (rotating wave approximation) are kept and we arrive at the following Hamiltonian
relevant for our discussion:
\begin{equation}
\label{eq:rwa}
H_{LI}^{RWA}=\frac{\hbar\Omega_{j}^{e}}{2}S_{j}^{+}
\left[e^{i(\Delta {\bf k}\cdot{\bf {\hat R}}_{j}(t)-\delta\omega t+\phi)}\right]
+H.C. ,
\end{equation}
in which the slow mode angular frequency is given by $\delta\omega =\Delta\omega-\omega_{0}$ and $\phi$ is the static phase shift between the laser beams.
The coupling of the reference Raman beam with photon frequency $\omega_{L}$ and the blue-detuned photon with frequency $\omega_{L}+\omega_{0}+\mu$ ($\mu>0$) in the second beam leads to an effective blue-detuned beam with the frequency difference $\Delta\omega=\omega_{0}+\mu$ as given by the Hamiltonian $H_{b}$
\begin{equation}
H_{b}=\hbar\sum_{j=1}^{N}\frac{\Omega_{j}^{e}}{2}(S_{j}^+e^{i(\Delta {\bf k}\cdot {\hat{{\bf R}}_{j}(t)}-\Delta\omega_{b}t+\phi_{b})}+H.c.),
\end{equation}
where $H_{b}$ is the interaction with the blue detuned ($\Delta\omega_{b}=\omega_{0}+\mu$) laser
 that has a beatnote frequency $\mu>0$ and $\Delta {\bf k}$ is the wavevector difference of the two interfering laser beams that generate the Raman coupling, $\hat{{\bf R}}_{j}(t)={\bf R}_{j}^{0}+\delta\hat{{\bf R}}_{j}(t)$ is the ion position operator with the equilibrium ion position ${\bf R}_{j}^{0}$ at site $j$.
Similarly, the coupling of the photon from the reference beam
with the photon in the red-detuned beam with frequency $\omega_{L}+\omega_{0}-\mu$ leads to the effective red-detuned
laser with the frequency difference $\Delta\omega=\omega_{0}-\mu$ given by the Hamiltonian $H_{r}$
\begin{equation}
H_{r}=\hbar\sum_{j=1}^{N}\frac{\Omega_{j}^{e}}{2}(S_{j}^+e^{i(\Delta {\bf k}\cdot {\hat{{\bf R}}_{j}(t)}-\Delta\omega_{r}t+\phi_{r})}+H.c.).
\end{equation}

Employing a superposition of multiple frequency components and adiabatically eliminating the dipole allowed excited states~\cite{PJLee} allows one to show that the interaction of laser beams with ions
consists of interactions between the reference beam $\omega_{L}$ and the other frequencies. As a result,
after the summations in Eqs.~(A4) and (A5), one arrives at the following expression:
\begin{widetext}
\begin{equation}
H_{LI}^{RWA}=H_{b}+H_{r}=\hbar\Omega_{j}^{e}\cos(\mu t+\phi_{M})\left[e^{i\Delta{\bf k}\cdot\delta {\bf \hat{R}}_{j}(t)}S_{j}^{+}e^{-i\phi_{S}}+e^{-i\Delta{\bf k}\cdot\delta {\bf {\hat R}}_{j}(t)}S_{j}^{-}e^{+i\phi_{S}}\right],
\end{equation}
 in which hermiticity of the Rabi frequency is used, and  the static phases are $\phi_{S}=-\Delta{\bf k}\cdot {\bf R}_{j}^{0}-(\phi_{b}+\phi_{r})/2$ and $\phi_{M}=(\phi_{r}-\phi_{b})/2$.
In the Lamb-Dicke limit, we have $\exp{[i\Delta{\bf k}\cdot\delta {\bf \hat{R}}_{j}]}\approx 1+i\Delta{\bf k}\cdot\delta {\bf \hat{R}}_{j}$, and
the Hamiltonian $H_{LI}^{RWA}$  is reduced to
\begin{equation}
H_{LI}^{RWA}\approx \hbar\Omega_{j}^{e}\cos(\mu t+\phi_{M})S_{j}^{+}e^{-i\phi_{S}}\left[
1+i\Delta{\bf k}\cdot\delta {\bf \hat{R}}_{j}(t)\right]+H.C.
\end{equation}
\end{widetext}
The first term only induces resonant carrier transitions in the pseudospin sector without coherent phonon excitations. The second term induces first (red or blue) side-band transitions with the change of one phonon occupation number at each phonon mode as can be seen by replacing the displacement operator $\delta{\bf \hat{R}}_{j}(t)$ by phonon creation and annihilation operators $\sum_{\alpha\nu}
 \frac{1}{\sqrt{\hbar/2M\omega_{\nu}}}b_{j}^{\alpha\nu}
 (a_{\alpha\nu}+a_{\alpha\nu}^{\dagger})$.
 The spin-dependent force pointing along the azimuthal angle $\phi_{S}^{'}$ in the equator of Bloch sphere  is then derived from the phonon side bands as
\begin{equation}
H_{LI}^{SB}\approx \hbar\Omega_{j}^{e}\cos(\mu t+\phi_{M})\sigma_{j}^{\phi_{S}^{'}}
\Delta{\bf k}\cdot\delta {\bf \hat{R}}_{j}(t),
\end{equation}
in which the spin phase is given by $\phi_{S}^{'}=\phi_{S}-\pi/2$,  the relation
$S_{j}^{+}e^{-i\phi_{S}^{'}}+S_{j}^{-}e^{+i\phi_{S}^{'}}=\sigma_{j}
^{\phi_{S}^{'}}$ is used, and the spin orientation is given by $\hat {\bf n}=\cos(\phi_{S}^{'}){\hat {\bf x}}+\sin(\phi_{S}^{'}){\hat {\bf y}}$.

The expression for the spin-dependent force can be justified
by keeping the phases $\phi_{S}^{'}=0,\phi_{M}=\pi/2$ locked. Take a transverse phonon mode scheme for example, in which
$\Delta{\bf k}\cdot{\bf R}_{j}^{0}=0$, with $\Delta{\bf k}\parallel\delta {\bf \hat{R}}_{j}(t)$.
The spin-orientation
$\phi_{S}^{'}=0$  can be locked along the $x$-axis in Bloch sphere when the phase difference $\phi_{r}=\pi/2,\phi_{b}=-\pi/2$ is maintained.
This can be achieved by passing the second beam through an acousto-optic modulator (AOM) maintaining the phase difference between the frequency components $\Omega_{L}+\omega_{0},\Omega_{L}+\omega_{0}+\mu,\Omega_{L}+\omega_{0}-\mu$
to be out of phase.
As one can tell from the dependence of the spin phase $\phi_{S}$ on $\Delta{\bf k}\cdot {\bf R}_{0}$, it is not sensitive to
transverse phonon excitations (coherent or thermal) in contrast to the sensitivity it has to the longitudinal phonon modes. This is why most state-of-art trapped ion quantum spin simulators couple to transverse phonon modes.

One should note that there is a fast oscillating term in the transverse magnetic field $-\hbar\Omega_j^e\sin(\mu t)\sigma_j^y$ due to carrier transitions. This term causes very fast oscillations of low amplitude which are averaged over during the time of an experiment, so we neglect them here.

Let us now consider how to generate a slow effective transverse magnetic field by using two continuous Raman beams with frequencies $\omega_{L}$, and $\omega_{L}+\omega_{0}$, with phase difference $\phi$, and wavevector difference $\Delta {\bf k}$.
Starting from Eq.~(A6) but with a different effective Rabi frequency $\Omega_{j}^{B}$ for the resonant beam with $\mu=0$
\begin{equation}
H_{LI}^{B}=\frac{\hbar\Omega_{j}^{B}}{2}S_{j}^{+}e^{i{\Delta{\bf k}\cdot [{{\bf R}_{j}^{0}+\delta\hat{\bf R}}_{j}(t)]+\phi}}+H.C.,
\end{equation}
we choose the lasers to be out of phase ($\phi=\mp\pi/2$) so that the side-band terms vanish within the Lamb-Dicke expansion $\exp{[i\Delta{\bf k}\cdot \delta{\bf R}_{j}(t)]}\approx 1+i\delta{\bf k}\cdot \delta{\bf R}_{j}(t)$ . The effective transverse magnetic field can then be derived by direct substitution as
\begin{equation}
H_{B}=B_{j}^{y}\sigma_{j}^{y},
\end{equation}
 in which the transverse magnetic field is given by $B_{j}^{y}=\pm \hbar\Omega_{j}^{B}/2$ when the phase shift $\phi$ is equal to $\mp\pi/2$ and $\sigma_{j}^{y}=2S_{j}^{y}$ is the Pauli spin operator (we will be working in a nontraditional Pauli spin matrix representation, where $\sigma^x$ is diagonal, $\sigma^y$ is real, and $\sigma^z$ is imaginary). Hence, the transverse magnetic field $B_{y}$ can have its amplitude changed as a function of time by adjusting the laser intensity in the mode that has its frequency equal to $\omega_{L}+\omega_{0}$  with an AOM.\\

\section{Adiabatic perturbation theory}
The time-dependent Schr\"odinger equation for the evolution of the wave function $|\psi(t)\rangle$ is (we drop the spin subscript on the Hamiltonian)
\begin{equation}
\label{eq:schrodinger}
i\hbar\frac{\partial |\psi(t)\rangle}{\partial t}= H(t)|\psi(t)\rangle.
\end{equation}
Since the Hamiltonian is always Hermitian, we introduce instantaneous eigenfunctions $|n(t)\rangle$ with the instantaneous eigenenergies defined by
\begin{equation}
\label{eq:instantneous}
H(t)|n(t)\rangle=E_{n}(t)|n(t)\rangle.
\end{equation}
The time-dependent wave function $|\psi(t)\rangle$ can then be expanded in terms of the orthonormal eigenbasis $|n(t)\rangle$
as
\begin{equation}
|\psi(t)\rangle = \sum_{n}C_{n}(t)|n(t)\rangle,
\end{equation}
in which the coefficients $C_{n}(t)=\langle n(t)|\psi(t)\rangle$ are the time-dependent quantum amplitudes projected onto the instantaneous eigenbasis $|n(t)\rangle$.
Therefore, the equation of motion for the expansion coefficients $C_{n}(t)$ can be derived by direct substitution into the Schr\"odinger equation in Eq.~(\ref{eq:schrodinger}) which becomes
\begin{equation}
\label{eq:coefficient}
i\hbar \Big[ \dot{C}_{m}(t)+\sum_{n}C_{n}(t)\langle m(t)|\partial_{t}|n(t)\rangle\Big]=E_{m}(t)C_{m}(t),
\end{equation}
after using the orthonormality relation $\langle m(t)|n(t)\rangle=\delta_{m,n}$ for the instantaneous eigenfunctions.
One can further relate the matrix elements $\langle m(t)|\partial_{t}|n(t)\rangle$ to the matrix elements $\langle m(t)|\partial_{t}H(t)|n(t)\rangle$. Simply take the time derivative of Eq.~(\ref{eq:instantneous})
and project onto the state $|m(t)\rangle$, to show
\begin{equation}
\label{eq:relation}
\langle m(t)|\partial_{t}|n(t)\rangle=\frac{\langle m(t)|\partial_{t}H(t)|n(t)\rangle}{E_{n}(t)-E_{m}(t)}
\end{equation}
for $n\ne m$ [this derivation assumes the instantaneous energy spectrum has no states that are degenerate with $E_n(t)$].

In the adiabatic approximation, the transition matrix
elements $\langle m(t)|\partial_{t}H(t)|n(t)\rangle$ between different instantaneous eigenstates are assumed to be so small they can be neglected.
In this limit, the system simply follows the instantaneous eigenstates $|n(t)\rangle$ without
transitions between different instantaneous eigenstates. In general, transitions between eigenstates $|n(t)\rangle$
should be considered to determine the corrections to the adiabatic state evolution.
Choose the gauge $C_{m}(t)=\alpha_{m}(t)e^{-i\theta_{m}(t)}$ with the phase $\theta_{m}(t)=\int_{t_{0}}^{t}dt^{\prime}\frac{E_{m}(t^{\prime})}{\hbar}$.
The probability amplitude $\alpha_{m}(t)$ induced by the diabatic transitions can be solved by integration with respect to time in Eq.~(\ref{eq:coefficient}) as
\begin{equation}
\label{eq:transitions}
\frac{d\alpha_{m}(t)}{dt}=\sum_{n} \frac{\langle m(t)|\partial_{t}H(t)|n(t)\rangle}{E_{m}(t)-E_{n}(t)}
e^{i[\theta_{m}(t)-\theta_{n}(t)]},
\end{equation}
in which $\theta_{m}(t)$ and $\theta_{n}(t)$ are the dynamic phases accumulated by the states $|m(t)\rangle$ and $|n(t)\rangle$.
%given respectively by
%\begin{equation}
%\theta_{m}(t^{\prime})=\int_{t_{0}}^{t^{\prime}}dt''\frac{E_{m}(t'')}{\hbar}, \theta_{n}
%(t^{\prime})=\int_{t_{0}}^{t^{\prime}}dt''\frac{E_{n}(t'')}{\hbar}.
%\end{equation}

In an adiabatic quantum simulation, one initially prepares the system in a certain pure state $|n(t_0)\rangle$ of the initial Hamiltonian $H(t_0)$ with the occupation $\alpha_{n}(t_{0})=1$ and the probability amplitudes in all other states vanishing [$\alpha_{m}(t_{0})=0$].
Therefore, the probability amplitudes to be excited into the other states can be approximated by
\begin{equation}
%\label{eq:excitations}
\alpha_{m}(t)\approx \int_{t_{0}}^{t}dt^{\prime} \frac{\langle m(t^{\prime})|\partial_{t^{\prime}}H(t^{\prime})|n(t^{\prime})\rangle}{E_{m}(t^{\prime})-E_{n}(t^{\prime})}e^{i[\theta_{m}(t^{\prime})-\theta_{n}(t^{\prime})]},
\end{equation}
for later times, as long as the transition amplitudes $|\alpha_{m}(t)|$ are much smaller than one during the time evolution.
This is the main expression we will use to evaluate the diabatic effects due to the time-dependent exchange interactions $J_{jj^{\prime}}(t)$.

\section{GHZ state entanglement}
The observable for the measurement of GHZ state entanglement is given by
\begin{equation}
\langle W_{\rm GHZ}\rangle=Tr\left\{\rho W_{\rm GHZ}\right\}
\end{equation}
with the density matrix $\rho$ constructed from either pure states or mixed states after tracing over the phonons. One can explicitly show that a pure GHZ state $|\rm GHZ\rangle=\frac{1}{\sqrt{2}}(|\uparrow_{x}\uparrow_{x}\cdots\uparrow_{x}\rangle
+|\downarrow_{x}\downarrow_{x}\cdots\downarrow_{x}\rangle)$
leads to an entanglement measure $\langle W_{\rm GHZ}\rangle$ to be $-1$.
The entanglement measure $\langle W_{\rm GHZ}\rangle$ is independent of the basis; therefore, it is natural to choose spin polarized states along the Ising $x$-axis in our discussion. Based on the GHZ state $|\rm GHZ\rangle$, the corresponding density
matrix $\rho_{\rm GHZ}=|\rm GHZ\rangle\langle\rm GHZ|$ in the Ising product state representation is given by
\begin{widetext}
\begin{equation}
\rho_{\rm GHZ}=\frac{1}{2}\Big[|\uparrow_{x}\uparrow_{x}\cdots\rangle
\langle\uparrow_{x}\uparrow_{x}\cdots|+|\downarrow_{x}\downarrow_{x}\cdots
\rangle
\langle\downarrow_{x}\downarrow_{x}\cdots |+
|\uparrow_{x}\uparrow_{x}\cdots \rangle
\langle\downarrow_{x}\downarrow_{x}\cdots |+
|\downarrow_{x}\downarrow_{x}\cdots\downarrow_{x}\rangle
\langle\uparrow_{x}\uparrow_{x}\cdots |
\Big].
\end{equation}
\end{widetext}
Therefore, the nonzero matrix elements in $\rho_{\rm GHZ}$ are among the fully ferromagnetic states and each has an expectation  value equal to $1/2$. As a consequence, one only needs to know the  $W_{\rm GHZ}$ matrix elements among the FM states to evaluate
the GHZ state entanglement $\langle W_{\rm GHZ}\rangle$. Following the definition of the witness operator $W_{\rm GHZ}$
\begin{equation}
W_{{\rm GHZ}}=3I-2\Big[\frac{S_{1}^{{\rm GHZ_{N}}}+I}{2}
+\prod_{k=2}^{N}\frac{S_{k}^{{\rm GHZ_{N}}}+I}{2}\Big]
\end{equation}
with the stabilizing spin operators $S_{1}^{{\rm GHZ_{N}}}, S_{k}^{{\rm GHZ_{N}}}$ expressed in terms of the Pauli spin operators by
\begin{equation}
S_{1}^{{\rm GHZ_{N}}}=\prod_{k=1}^{N}\sigma^{y}_{k},\quad
S_{k}^{{\rm GHZ_{N}}}=\sigma^{x}_{k-1}\sigma^{x}_{k},
\end{equation}
the nonzero matrix elements for the global spin flipping operator $S_{1}^{\rm GHZ_{N}}$ amongst the $N$ ion FM states is
\begin{widetext}
\begin{equation}
\langle \uparrow_{x}\uparrow_{x}\cdots|S_{1}^{\rm GHZ_{N}}|\downarrow_{x}\downarrow_{x}\cdots\rangle=
\langle\downarrow_{x}\downarrow_{x}\cdots|S_{1}^{\rm GHZ_{N}}|\uparrow_{x}\uparrow_{x}
\cdots\rangle
=1,
\end{equation}
\end{widetext}
in which the spin operator $\sigma^{y}_{i}$ flips the spin state of $i$-th
ion as $\sigma^{y}_{i}|\uparrow_{x}\rangle =|\downarrow_{x}\rangle$ and $\sigma^{y}_{i}|\downarrow_{x}>\rangle=|\uparrow_{x}\rangle$ with $I$ the unit operator for spin states.

One can also derive the nonzero matrix elements for the  two-particle spin
operator $S_{k}^{GHZ_{N}}$ as:
\begin{widetext}
\begin{equation}
\langle \uparrow_{x}\uparrow_{x}\cdots| \prod_{k=2}^{N}{S_{k}^{\rm GHZ_{N}}+I}|\downarrow_{x}\downarrow_{x}\cdots\rangle
=\langle \downarrow_{x}\downarrow_{x}\cdots| \prod_{k=2}^{N}{S_{k}^{\rm GHZ_{N}}+I}|\uparrow_{x}\uparrow_{x}\cdots\rangle
=2^{N-1}.
\end{equation}
\end{widetext}
As a consequence, the only nonzero matrix elements for the witness operator $W_{\rm GHZ}$ are the ones between two FM states which become
\begin{equation}
\begin{array}{lll}
  \langle \uparrow_{x}\uparrow_{x}\cdots|W_{\rm GHZ}|\downarrow_{x}\downarrow_{x}
\cdots\rangle   &  \\
 =\langle\downarrow_{x}\downarrow_{x}\cdots|W_{\rm GHZ}|\uparrow_{x}\uparrow_{x}
\cdots\rangle &  \\
= -1.&
\end{array}
\end{equation}
Hence, the measurement $\langle W_{\rm GHZ}\rangle$ for the  pure $\rm GHZ$ state is characterized  by the value $-1$ as can be seen by the following manipulations
\begin{equation}
\begin{array}{ll}
   \langle W_{\rm GHZ}\rangle=\langle \uparrow_{x}\cdots|\rho_{\rm GHZ}|\downarrow_{x}
\cdots\rangle\langle \downarrow_{x}\cdots|W_{\rm GHZ}|\uparrow_{x}
\cdots\rangle & \\
  +
\langle \downarrow_{x}\cdots|\rho_{\rm GHZ}|\uparrow_{x}
\cdots\rangle\langle \uparrow_{x}\cdots|W_{\rm GHZ}|\downarrow_{x}
\cdots\rangle &  \\
  =\frac{1}{2}\times(-1)+\frac{1}{2}\times(-1) =-1. &
\end{array}
\end{equation}

\end{document}